\documentclass[aps,prd,reprint,groupedaddress]{revtex4-2}
\usepackage{graphicx}
\usepackage{dcolumn}
\usepackage{bm}
\usepackage{hyperref}
\usepackage[utf8]{inputenc}
\usepackage{color}
\usepackage{amsmath}
\allowdisplaybreaks[4]
\numberwithin{equation}{section}
\usepackage{amssymb}

\newcommand{\mc}{\mathcal}
\newcommand{\mb}{\mathbf}

\def \be {\begin{equation}}
\def \ee {\end{equation}}
\def \ba {\begin{array}}
\def \ea {\end{array}}
\def \bea{\begin{eqnarray}}
\def \eea{\end{eqnarray}}
\newcommand{\SOUTHCUT}{\vspace{0.5em} School of Physics and Optoelectronics, South China University of Technology, \\ Guangzhou 510641, People's Republic of China}

\begin{document}

%\preprint{}

\title{Angular momentum flux through post-Newtonian order for noncircular nonspinning black-hole binaries in Einstein-Maxwell-dilaton theory}


\author{ Qi Tan}
%\thanks{}
 \email{202230351161@mail.scut.edu.cn}
 \author{Peng-Cheng Li}
 \email{pchli2021@scut.edu.cn (corresponding author)}
\affiliation{\SOUTHCUT}


\date{\today}

\begin{abstract}
We compute the instantaneous angular momentum flux from nonspinning black hole binaries in Einstein-Maxwell-dilaton theory for generic noncircular orbits up to relative first post-Newtonian order. Working in the Einstein frame and using the direct integration of the relaxed field equations approach, we construct the scalar, electromagnetic, and tensor gravitational fields in the wave zone and express the required source multipole moments in the center-of-mass frame. In addition to the leading $1/R$ radiative fields, we retain the next-to-leading $1/R^2$ terms in the wave-zone fields, which describe finite-distance corrections and do not contribute to the flux at null infinity. We obtain separately the scalar, electromagnetic, and tensor contributions to the angular momentum flux. The scalar and electromagnetic channels begin with dipole radiation, while the tensor channel begins at quadrupole order. Our results recover the quasicircular balance relation, the known general-relativistic limit, and the expected scalar- and electromagnetic-dipole suppression limits. Together with the previously known energy flux, these results provide the dissipative information required for future studies of orbit-averaged eccentric evolution and waveform phasing in Einstein--Maxwell--dilaton theory.
\end{abstract}

%\keywords{}

\maketitle

\section{Introduction}
The first direct detection of gravitational waves from a binary black-hole merger and the rapidly expanding catalog of compact-binary coalescences have established gravitational-wave astronomy as a powerful probe of relativistic gravity \cite{LIGOScientific:2016aoc,LIGOScientific:2026wfs}. In addition to revealing the astrophysical properties of compact objects, gravitational-wave observations provide access to the nonlinear and dynamical strong-field regime, in which deviations from general relativity (GR) may become observable \cite{Yunes:2013dva}. Extracting such information requires accurate theoretical descriptions of both the conservative binary dynamics and the emitted radiation. In the weak-field and slow-motion stages of the inspiral, the post-Newtonian (PN) approximation provides a systematic framework for constructing these descriptions \cite{Blanchet:2013haa}.

Many theories beyond vacuum GR contain additional scalar or vector degrees of freedom. Such fields can modify the structure of compact objects, introduce additional radiation channels, and alter the orbital evolution of binary systems \cite{Henry:2023guc, Henry:2023len, Bhattacharyya:2023kbh,Ghosh:2025ban}. Although black holes are subject to powerful no-hair results under specific assumptions, these assumptions can be evaded when additional fields are coupled nonminimally to gravity or to one another \cite{Sotiriou:2013qea,Herdeiro:2014goa,Herdeiro:2015waa,Cardoso:2016ryw}. Einstein--Maxwell--dilaton (EMd) theory provides a particularly useful setting in which one can investigate these effects. It arises in low-energy limits of string-inspired theories and contains a scalar dilaton coupled nonminimally to the electromagnetic field. Electrically charged black holes in EMd theory can support a nontrivial dilaton profile and therefore carry scalar hair in addition to their mass and electromagnetic charge \cite{GIBBONS1988741,PhysRevD.43.3140} (see \cite{Herdeiro:2018wub,Fernandes:2019rez,Blazquez-Salcedo:2020nhs,Hod:2022txa,Lai:2022spn,Lai:2022ppn,Guo:2023mda,Xiong:2023bpl,Herdeiro:2025blx} for developments in more general Einstein–Maxwell–scalar theory). Although astrophysical black holes are generally expected to carry only small electric charges in conventional scenarios \cite{Gibbons:1975kk,Blandford:1977ds}, EMd theory serves as a tractable theoretical laboratory for studying binaries with coupled tensor, scalar, and vector degrees of freedom \cite{Hirschmann:2017psw,Khalil:2018aaj, Bhattacharyya:2025ghl}.

The effective point-particle description of hairy black holes in EMd theory was developed in Ref.~\cite{Julie:2017rpw}, where the scalar-field dependence of the effective masses was used to incorporate the strong-field properties of the individual objects. The conservative and dissipative dynamics of EMd black-hole binaries were subsequently investigated within the PN and effective-one-body frameworks in Ref.~\cite{Khalil:2018aaj}. The energy carried by the tensor, scalar, and electromagnetic radiation channels was also computed for circular binaries at leading multipolar order in Ref.~\cite{Julie:2018lfp}. These studies provide the essential ingredients for modeling quasicircular EMd binaries. A corresponding calculation of the angular momentum flux for generic noncircular motion, however, is still required in order to describe the evolution of binaries with an independent radial degree of freedom.

This distinction is important because the dissipative evolution of a generic noncircular orbit cannot be determined from the energy flux alone. For a quasicircular binary, the orbital energy and angular momentum can both be parametrized by a single orbital frequency, and their fluxes are related by the circular-orbit balance relation. For a noncircular binary, by contrast, the orbital energy and angular momentum are independent quantities. Their secular evolution therefore requires two independent balance equations involving both the energy and angular momentum fluxes \cite{DeFalco:2023djo, DeFalco:2024ojf}. This is the basis of the classic treatment of eccentric inspirals in GR \cite{Peters:1963ux,Peters:1964zz}, as well as higher-PN calculations of the angular momentum flux and orbital-element evolution for quasielliptical compact binaries \cite{Arun:2009mc}. Similar energy and angular-momentum balance calculations have recently been carried out for eccentric binaries in scalar--tensor theories \cite{Trestini:2024zpi, AbhishekChowdhuri:2022ora}. The EMd problem is distinguished by the simultaneous presence of scalar, electromagnetic, and tensor radiation and by the direct coupling between the scalar and electromagnetic fields.

Noncircular motion is also astrophysically relevant. Although gravitational radiation generally drives an isolated binary toward circularity, binaries assembled through dynamical encounters in dense stellar environments can retain appreciable eccentricity when entering the observational frequency band \cite{Samsing:2017xmd}. The additional dissipative channels present in EMd theory can, in principle, further modify the evolution of the orbital energy, angular momentum, and eccentricity. Determining these effects requires the angular-momentum loss for generic motion rather than a result restricted to circular trajectories.

In this paper, we compute the instantaneous angular momentum flux generated by a nonspinning EMd black-hole binary on a generic noncircular orbit through relative first post-Newtonian order. We derive separately the scalar, electromagnetic, and tensor gravitational contributions and combine them to obtain the total flux. Together with the previously known energy flux \cite{Khalil:2018aaj}, our results provide the dissipative information required for future calculations of orbit-averaged eccentric evolution and gravitational-wave phasing in EMd theory.

We also construct the scalar, electromagnetic, and gravitational fields in the wave zone. The fields are expanded in inverse powers of the distance \(R\) from the binary. Their leading $\mathcal{O}(R^{-1})$ components constitute the radiative fields and determine the angular momentum flux at null infinity. We additionally retain the next-to-leading $\mathcal{O}(R^{-2})$ terms as finite-distance corrections to the wave-zone fields. These terms do not contribute to the flux at null infinity and should therefore be regarded as a separate extension of the waveform calculation rather than as ingredients required for the angular momentum flux.

Our calculation is performed in the Einstein frame using the direct integration of the relaxed field equations (DIRE) approach \cite{1975ApJ...197..717E,Will:1996zj, Pati:2000vt, Pati:2002ux}. We first rewrite the coupled gravitational, scalar, and electromagnetic field equations as flat-spacetime wave equations with nonlinear effective sources. The retarded solutions are divided into near-zone and wave-zone contributions. The near-zone contributions are then organized in terms of scalar, electromagnetic, and Epstein--Wagoner gravitational source moments. After evaluating the required moments through the appropriate PN orders, we transform them to the center-of-mass frame and construct the corresponding wave-zone fields. Finally, the radiative $\mathcal{O}(R^{-1})$ fields are inserted into the scalar, electromagnetic, and gravitational angular momentum flux formulas, yielding the instantaneous total flux through relative 1PN order.

The remainder of this paper is organized as follows. In Sec.~\ref{sec:model}, we introduce the EMd action, the effective point-particle description of the compact objects, and the relaxed gravitational, scalar, and electromagnetic field equations. In Sec.~\ref{sec:formal_solution}, we describe the DIRE construction, separate the near-zone and wave-zone contributions, and derive the source multipole moments and the inverse-distance expansions of the wave-zone fields. In Sec.~\ref{sec:CM_Waveforms}, we reduce the binary dynamics and the multipole moments to the center-of-mass frame and construct the scalar, electromagnetic, and tensor fields at the PN orders required in this work. In Sec.~\ref{sec:flux_evo}, we derive the conservative orbital angular momentum and compute the scalar, electromagnetic, gravitational, and total instantaneous angular momentum fluxes through relative 1PN order. We summarize our results and discuss possible extensions in Sec.~\ref{sec:conclu}.

\section{The Model and Relaxed Field Equations} \label{sec:model}

The action of EMd theory in the Einstein frame  is expressed as \cite{Khalil:2018aaj, Astefanesei:2019pfq}
\begin{align} \label{Def_action}
    S &= \frac{c^3}{16\pi G} \int d^4x \sqrt{-g} \bigg[ R - 2g^{\mu\nu}\partial_\mu\varphi\partial_\nu\varphi -\frac{G}{c^4} f(\varphi)F_{\mu\nu}F^{\mu\nu} \bigg] \notag \\
    &\qquad + S_m (\mathcal{A}^2 (\varphi) g_{\mu \nu}, A_\mu, \psi).
\end{align}
Here, $g$ denotes the determinant of metric $g_{\mu \nu}$, $R$ is the Ricci scalar, defined by metric $g_{\mu \nu}$. $\varphi$ is the scalar field (dilaton), $F_{\mu\nu} = \partial_\mu A_\nu - \partial_\nu A_\mu$ is the electromagnetic field tensor, and we adopt the dilaton coupling $f(\varphi) = e^{-2a\varphi}$ later. For the matter part $S_m$, $\mathcal{A} (\varphi) = e^{a \varphi}$ represents the conformal transformation from the Jordan frame into the Einstein frame (For the explanation of equivalence between the two frames, see Ref. \cite{Khalil:2018aaj}). $A_\mu$ is electromagnetic four-potential, and $\psi$ is the pure matter contribution to the matter action.

For the explicit form of matter action, we model the components of many-body system as point particles. This effective matter action is constructed by Eardley in the context of Brans-Dicke theory. The matter action $S_m$ is given by \cite{Eardley:1975fgi}
\begin{equation}\label{pointparticleaction}
    S_m = -\sum_A \int dt \left[ \mathfrak{m}_A(\varphi) c \sqrt{-g_{\mu\nu}v^\mu_A v^\nu_A} - \frac{1}{c}q_A A_\mu v^\mu_A \right],
\end{equation}
where $v^\mu_A = (c, v^j_A)$ is the coordinate four-velocity of particle $A$, and $q_A$ is its electric charge. The scalar-dependent mass $\mathfrak{m}_A(\varphi)$ implies that the scalar field directly couples to matter.

In general, $\mathfrak{m}_A (\varphi)$ does not have analytic closed-form expression. In the PN near zone, the scalar-field perturbation generated by the companion scales as $\delta\varphi \sim GM/(rc^2)=\mc{O}(c^{-2})$. The scalar-dependent mass can therefore be Taylor-expanded about the asymptotic background value $\varphi_0$ as
\begin{equation} \label{Def_mass}
    \mathfrak{m}(\varphi) = m \left[ 1 + \alpha \delta \varphi + \frac{1}{2} (\alpha^2 + \beta) \delta \varphi^2 + \mathcal{O}(c^{-6}) \right],
\end{equation}
where $m$ is the mass in the background scalar field $\varphi_0$ and parameters $\alpha$ and $\beta$ are defined by
\begin{equation}
    \alpha(\varphi) = \frac{d \ln \mathfrak{m}(\varphi)}{d\varphi}, \quad \beta(\varphi) = \frac{d^2 \ln \mathfrak{m}}{d\varphi^2}.
\end{equation}
Here, $\alpha$ is the dimensionless scalar charge, an analogue of electric charge in the scalar-tensor theory. Notice that $\alpha(\varphi)$ is scalar-field-dependent, and we adopt simplified notations $\alpha = \alpha(\varphi_0)$ and $\beta = \beta(\varphi_0)$ in Eq.~\eqref{Def_mass} and in later sections of this paper.

The scalar-field dependence of the effective scalar charge \(\alpha(\varphi)\) has been studied in Refs. \cite{Khalil:2018aaj,Julie:2017rpw}. In \cite{Julie:2017rpw}, Juli\'e expanded both the Garfinkle--Horowitz--Strominger  black-hole solution \cite{PhysRevD.43.3140} and the solution generated by the skeletonized point-particle model in the asymptotic region. Matching the two results through $\mathcal{O}(R^{-1})$, yields a differential equation for the scalar-dependent mass $\mathfrak{m}(\varphi)$, and hence determines the scalar charge $\alpha(\varphi)$.

\subsection{Relaxed Field Equations}
By varying the action ($ \delta S = 0 $) with respect to $g^{\mu \nu}$, $\varphi$ and $A_{\nu}$, we obtain the coupled field equations \cite{Astefanesei:2019pfq}.

First, the gravitational field equation is given by
\begin{equation}
    G_{\mu\nu} = \frac{8\pi G}{c^4} \left( T_{\mu\nu}^\varphi + T^A_{\mu\nu} + T_{\mu\nu} \right), \label{eq:grav_eq}
\end{equation}
where the stress-energy tensors for the scalar field, the electromagnetic field, and the matter source are respectively
\begin{align}
    T_{\mu\nu}^\varphi &= \frac{c^4}{4\pi G} \left( \partial_\mu\varphi\partial_\nu\varphi - \frac{1}{2}g_{\mu\nu}\partial_\rho\varphi\partial^\rho\varphi \right), \\
    T^A_{\mu\nu} &= \frac{1}{4\pi} e^{-2a\varphi} \left( F_{\mu\rho}F_\nu^{\ \rho} - \frac{1}{4}g_{\mu\nu}F_{\rho\sigma}F^{\rho\sigma} \right), \\
    T_{\mu\nu} &= -\frac{2}{\sqrt{-g}} \frac{\delta\mathcal{L}_m}{\delta g^{\mu\nu}}.\label{matterenergymomentnsor}
\end{align}

Second, the equation of motion for the scalar field is
\begin{equation}
    \square_g \varphi = \frac{4\pi G}{c^4}(T^A + T^\varphi), \label{eq:scalar_eq}
\end{equation}
where $\square_g = g^{\mu\nu}\nabla_\mu\nabla_\nu$ is the d'Alembertian operator in the curved spacetime, and the source terms are defined as
\begin{align}
    T^A &= -\frac{a}{8\pi} e^{-2a\varphi} F_{\rho\sigma}F^{\rho\sigma}, \\
    T^\varphi &= -\frac{1}{\sqrt{-g}}\frac{\delta\mathcal{L}_m}{\delta\varphi}.\label{Tvarphi}
\end{align}

The electromagnetic field equation is
\begin{equation}
    \nabla_\mu \left[ f(\varphi)F^{\mu\nu} \right] = -4\pi j^\nu, \label{eq:em_eq}
\end{equation}
where the four-current $j^\nu$ is given by:
\begin{equation}
    j^\nu = \frac{1}{\sqrt{-g}}\frac{\delta\mathcal{L}_m}{\delta A_\nu}.\label{jnu}
\end{equation}

To solve these strongly coupled equations iteratively in the post-Newtonian approximation, we employ the Landau-Lifshitz formulation to transform them into a relaxed form \cite{Poisson_Will_2014, Will:1996zj, Landau:1975pou}. We define the gothic metric variable $\mathfrak{g}^{\mu\nu} = \sqrt{-g}g^{\mu\nu}$ and introduce the metric deviation
\begin{equation}
    h^{\mu\nu} = \eta^{\mu\nu} - \mathfrak{g}^{\mu\nu}.
\end{equation}

Imposing the harmonic coordinate condition $\partial_\mu \mathfrak{g}^{\mu\nu} = 0$, which is equivalent to $\partial_\mu h^{\mu\nu} = 0$, the exact gravitational equation \eqref{eq:grav_eq} reduces to a formal wave equation in the flat spacetime
\begin{equation}
    \square h^{\mu\nu} = -\frac{16\pi G}{c^4} \mu^{\mu\nu},
\end{equation}
where $\square = \eta^{\rho\sigma}\partial_\rho\partial_\sigma$ is the flat d'Alembertian operator. The effective source $\mu^{\mu\nu}$ incorporates the following terms
\begin{equation} \label{Def_gravitational_eff_source}
    \mu^{\mu\nu} = (-g)(T_\varphi^{\mu\nu} + T_A^{\mu\nu} + T^{\mu\nu} + t_{\text{LL}}^{\mu\nu} + t_H^{\mu\nu}),
\end{equation}
where $t_{\text{LL}}^{\mu\nu}$ is the Landau-Lifshitz pseudotensor and $t_H^{\mu\nu}$ is the harmonic gauge term (see Ref. \cite{Poisson_Will_2014} for explicit expression).

Similarly, we introduce a perturbation expansion for the scalar field
\begin{equation}
    \delta\varphi = \varphi - \varphi_0,
\end{equation}
where $\varphi_0$ is the asymptotic background value. By setting the asymptotic background to zero ($\varphi_0 = 0$), the relaxed field equation of scalar field is equivalent to
\begin{equation}
    \square \varphi = -\frac{4\pi G}{c^4}\mu,
\end{equation}
where the effective scalar source $\mu$ is
\begin{equation} \label{Def_scalar_eff_source}
    \mu = -\sqrt{-g}(T^A + T^\varphi) - \frac{c^4}{4\pi G}h^{\mu\nu}\partial_\mu\partial_\nu\varphi.
\end{equation}

Finally, for the electromagnetic field, we adopt the Lorenz gauge $\partial_\mu A^\mu = 0$. The relaxed equation for the vector potential $A^\nu$ is
\begin{equation}
    \square A^\nu = -4\pi \mu^\nu,
\end{equation}
where the effective current source $\mu^\nu$ is defined as
\begin{equation}
    \begin{split}
            \mu^\nu = & \sqrt{-g} \left( e^{2a\varphi} j^\nu - \frac{a}{2\pi} \partial_\mu \varphi F^{\mu\nu} \right) \\
& - \frac{1}{4\pi}(h^{\mu\rho}\partial_\mu\partial_\rho A^\nu - \partial_\mu h^{\nu\rho}\partial_\rho A^\mu) \\
& + \frac{1}{4\pi} \partial_\alpha(-\mathfrak{g}^{\alpha\mu}A_\sigma\partial_\mu g^{\nu\sigma}+\mathfrak{g}^{\nu\sigma}A_\mu\partial_\sigma g^{\alpha\mu}).
    \end{split}
\end{equation}

\subsection{Matter Source}
Using the definitions in Eqs.~(\ref{matterenergymomentnsor}), (\ref{Tvarphi}), and (\ref{jnu}), together with the point-particle action (\ref{pointparticleaction}) and the expansion of the scalar-dependent mass in Eq.~(\ref{Def_mass}), we now evaluate the matter stress-energy tensor, the scalar matter source, and the electromagnetic current. For convenience, we introduce the generalized Lorentz factor
\begin{equation}
    \gamma = \frac{c}{\sqrt{-g_{\mu\nu}v^\mu v^\nu}}.
\end{equation}
Substituting the point-particle action and the expansion ~\eqref{Def_mass} into the corresponding functional derivatives, we obtain
\begin{equation}
    T^{\mu\nu} = \frac{\rho_g \gamma v^\mu v^\nu}{\sqrt{-g}} \left[ 1 + \alpha\varphi + \frac{1}{2}(\alpha^2+\beta)\varphi^2 + \mathcal{O}(c^{-6}) \right],
\end{equation}
\begin{equation}
    T^\varphi = \frac{\rho_g c^2}{\gamma \sqrt{-g}} \left[ \alpha + (\alpha^2+\beta)\varphi + \mathcal{O}(c^{-4}) \right],
\end{equation}
and the four-current:
\begin{equation}
    j^\nu = \frac{\rho_e v^\nu}{c \sqrt{-g}},
\end{equation}
where $\rho_g$ and $\rho_e$ are respectively the mass density and electric charge density defined by
\begin{align}\label{rhogandrhoe}
    \rho_g &= \sum_A m_A \delta^3(\mathbf{x} - \mathbf{x}_A), \quad \rho_e = \sum_A q_A \delta^3(\mathbf{x} - \mathbf{x}_A).
\end{align}

\section{Formal solution for relaxed field equations} \label{sec:formal_solution}

\subsection{Wave-zone field point and the shortwave approximation}

Formal solutions to the relaxed field equations for the metric perturbation $h^{\mu\nu}$, the scalar field $\varphi$, and the vector potential $A^{\nu}$ are obtained across all regions of spacetime by employing the appropriate retarded Green's function \cite{Shiralilou:2021mfl, Poisson_Will_2014}. This yields the following integral expressions:
\begin{align}
    h^{\mu\nu}(x) &= \frac{4G}{c^{4}}\int\frac{\mu^{\mu\nu}(t', \mathbf{x}')\delta(t' - t + |\mathbf{x}-\mathbf{x}'|/c)}{|\mathbf{x}-\mathbf{x}'|}d^{4}x', \label{eq:gravitational_formal_solution} \\
    \varphi(x) &= \frac{G}{c^{4}}\int\frac{\mu(t', \mathbf{x}')\delta(t' - t + |\mathbf{x}-\mathbf{x}'|/c)}{|\mathbf{x}-\mathbf{x}'|}d^{4}x', \label{eq:scalar_formal_solution} \\ 
    A^{\nu}(x) &= \int\frac{\mu^{\nu}(t', \mathbf{x}')\delta(t' - t + |\mathbf{x}-\mathbf{x}'|/c)}{|\mathbf{x}-\mathbf{x}'|}d^{4}x', \label{eq:electromagnetic_formal_solution}
\end{align}
where the integration is performed over the past null cone of the field point $x=(ct, \mathbf{x})$ and $\delta(...)$ is the Dirac $\delta$ function. The evaluation of these integrals depends on the distance $R = |\mathbf{x}|$ compared to the characteristic wavelength $\lambda_c \sim c r_c/v_c$ of the radiation from a source of size $r_c$ and its characteristic speed $v_c$.

Now, we discuss the counting of PN order briefly. In this paper, we refer to 1PN order as an abbreviation of ``relative" 1PN order, i.e., since the scalar matter source term $\mu = \mathcal{O}(c^2)$, the scalar field $\varphi$ is of order $\mathcal{O}(c^{-2})$ (see Eq.~\eqref{eq:scalar_formal_solution}), then (relative) 1PN order contributions to the scalar field is of order $\mathcal{O}(c^{-4})$. However, for the gravitational field, the physical waveform information is contained in the spatial components $h^{jk}$, which are of order $\mathcal{O}(c^{-4})$, then 1PN order corrections for the gravitational field are of order $\mathcal{O}(c^{-6})$. In later sections, we omit the word ``relative" without leading to misunderstanding.

To evaluate the waveforms up to $\mathcal{O}(R^{-2})$ and the radiative angular momentum flux of the binary BH system, we examine the behavior of the fields in the far-away wave zone, where $R \gg \lambda_c$. Following the approach of Misner, Thorne, and Wheeler (1973) \cite{Misner:1973prb}, we employ the shortwave approximation \cite{Misner:1973prb, Khalil:2018aaj, Shiralilou:2021mfl, Poisson_Will_2014}, which utilizes $\lambda_c/R \ll 1$ as a formal expansion parameter. In this region, we expand the field $f$ in powers of $1/R$:
\begin{equation}
    f(t, \mathbf{x}) = \frac{\lambda_c}{R} f_1(\tau, \mathbf{N}) + \left( \frac{\lambda_c}{R} \right)^2 f_2(\tau, \mathbf{N}) + \dots, \label{eq:shortwave_expansion}
\end{equation}
where $\tau = t - R/c$ is the retarded time and $\mathbf{N} = \mathbf{x}/R$ is the unit direction vector from the coordinate origin, chosen at the center of mass of the binary, toward the wave-zone field point (or detector). In the far-zone limit, \(\mathbf{N}\) also represents the leading propagation direction from the source region to the observer.

\begin{figure*}[t]
    \centering
    \includegraphics[width=0.85\textwidth]{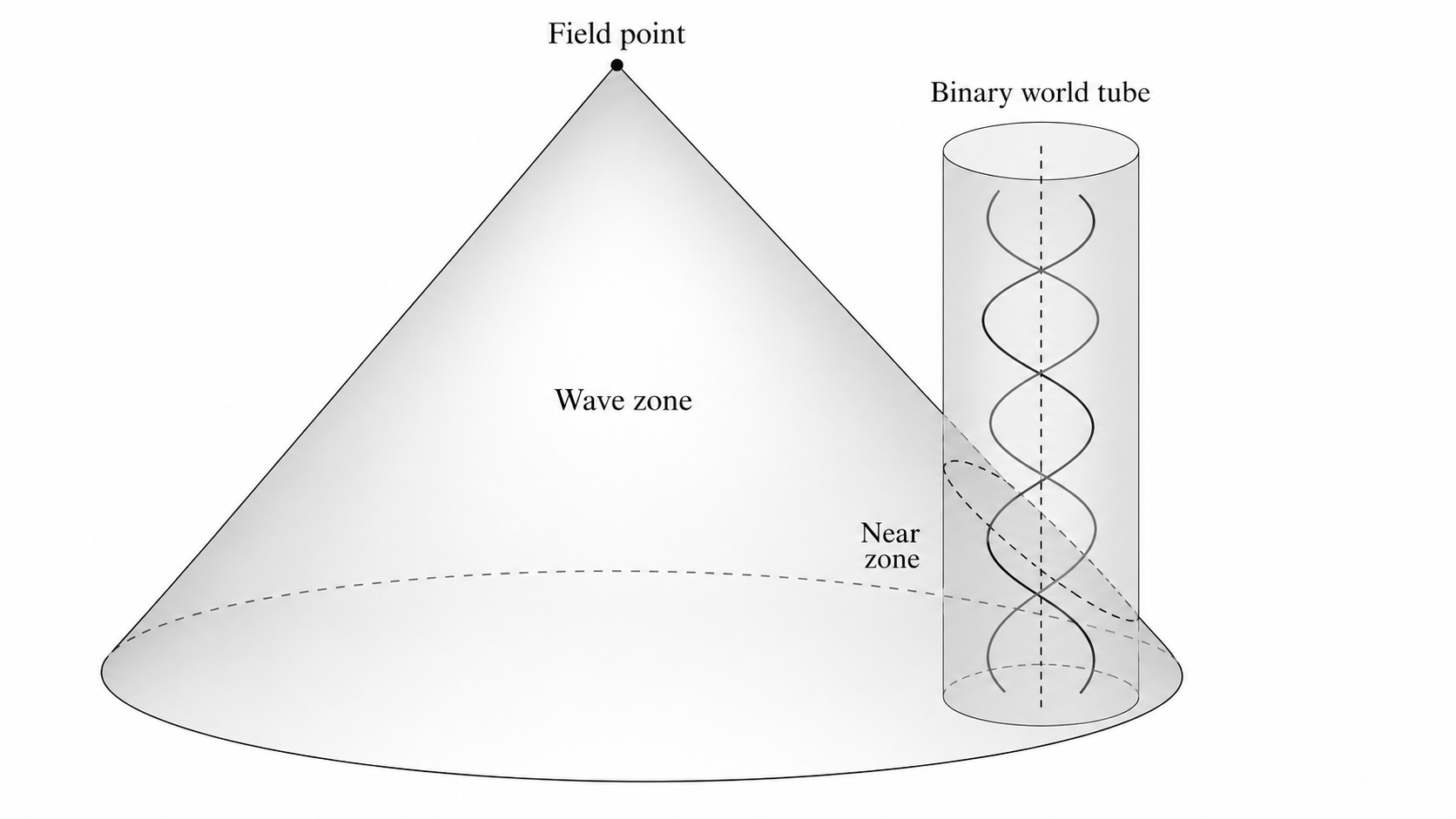}
    \caption{
        Schematic illustration of the DIRE decomposition for a field point
        located in the far zone. The past null cone of the field point
        intersects the near-zone world tube centered on the binary
        center-of-mass worldline. The retarded integral is divided into
        contributions from the near zone and the wave zone.
    }
    \label{fig:dire-schematic}
\end{figure*}

Following the DIRE approach, the retarded integrals are split into contributions from two spatial regions (see Fig.~\ref{fig:dire-schematic}): the near zone \(\mathcal{M}\), which contains the compact binary and is bounded by an arbitrary radius \(\mathcal{R}\), and the wave zone \(\mathcal{W}\), which lies outside this boundary. This split should be understood as a decomposition of the effective sources appearing in the relaxed field equations, rather than as a decomposition of the matter source itself. Although the point-particle matter sources are confined to the near zone, the effective sources \(\mu^{\mu\nu}\), \(\mu\), and \(\mu^\nu\) also contain nonlinear field contributions and are therefore not compactly supported. The total field at a wave-zone field point is thus obtained by adding the near-zone and wave-zone parts of the retarded integral. Within the PN expansion, these effective sources are expanded perturbatively in powers of \(1/c^2\), and the dependence on the artificial matching radius \(\mathcal{R}\) between the two regions cancels in the final waveform \cite{Will:1996zj}. To understand this cancellation, one can consider any integral over the all spacetime $f_{\text{all spacetime}}=f_\mathcal{M}(\mathcal{R})+f_\mathcal{W}(\mathcal{R})$ which contain integral contributions from both the near zone $\mathcal{M}$ and the wave zone $\mathcal{W}$. Since the integral $f_{\text{all spacetime}}$ depends on the volume of all spacetime only, instead of how we split this volume, although the integral in the near zone $\mathcal{M}$ (or the wave zone $\mathcal{W}$) indeed depends on how we split the volume, the sum of integral over the near zone and that over the wave zone cannot depend on the artificial matching radius $\mathcal{R}$. Thus, for convenience, we omit the $\mathcal{R}$-dependent terms appeared in the near zone or the wave zone contribution, even we consider them separately.

For the wave-zone field solutions considered in this work, the near-zone contribution gives the usual multipolar expansion of the radiative field. The wave-zone contribution, on the other hand, comes from nonlinear effective sources distributed outside \(\mathcal{M}\). As shown in Appendix \ref{app:wavezone_contribution}, these wave-zone source contributions start at order \(R^{-2}\) in the far-zone expansion. They therefore represent finite-distance corrections to the waveform, rather than corrections to the leading radiative field. Since the angular momentum flux at null infinity is determined by the \(\mc{O}(R^{-1})\) radiative pieces of the scalar, electromagnetic, and gravitational fields, the \(\mc{O}(R^{-2})\) wave-zone contributions vanish after taking the limit \(R\to\infty\) in the flux formula and do not affect the angular momentum flux computed below.

\subsection{Near-zone field solutions} \label{sec:sol_nearzone}

To calculate the 1PN waveforms to $\mathcal{O}(R^{-2})$ under the shortwave approximation, it is necessary to determine the 1PN expansion of the effective source $\mu$ within the near-zone. Khalil et al. \cite{Khalil:2018aaj} derived the 1PN solutions for the gravitational, scalar, and electromagnetic fields in the near-zone using the Fokker action method \cite{Fokker1929} (see also \cite{Damour:1995kt, Bernard:2015njp, Bernard:2018hta} for references):
\begin{widetext}
    \begin{equation}
        \begin{split}
            V &= \frac{G}{c^2} \left( \frac{m_1}{|\mathbf{x}-\mathbf{x}_1|} + \frac{m_2}{|\mathbf{x}-\mathbf{x}_2|} \right) + \frac{3G}{2c^4} \left( \frac{m_1 v_1^2}{|\mathbf{x}-\mathbf{x}_1|} + \frac{m_2 v_2^2}{|\mathbf{x}-\mathbf{x}_2|} \right) + \frac{G}{2c^4} m_1 \left( \frac{v_1^2}{|\mathbf{x}-\mathbf{x}_1|} - \mathbf{n}_1 \cdot \mathbf{a}_1 - \frac{(\mathbf{n}_1 \cdot \mathbf{v}_1)^2}{|\mathbf{x}-\mathbf{x}_1|} \right) \\
            &+ \frac{G}{2c^4} m_2 \left( \frac{v_2^2}{|\mathbf{x}-\mathbf{x}_2|} - \mathbf{n}_2 \cdot \mathbf{a}_2 - \frac{(\mathbf{n}_2 \cdot \mathbf{v}_2)^2}{|\mathbf{x}-\mathbf{x}_2|} \right) - \frac{G^2}{c^4} \left( \frac{m_1 m_2 \alpha_1 \alpha_2}{r |\mathbf{x}-\mathbf{x}_1|} + \frac{m_1 m_2 \alpha_1 \alpha_2}{r |\mathbf{x}-\mathbf{x}_2|} \right) \\
            &- \frac{G^2}{c^4} \left( \frac{m_1 m_2}{r |\mathbf{x}-\mathbf{x}_1|} + \frac{m_1 m_2}{r |\mathbf{x}-\mathbf{x}_2|} \right) - \frac{G}{2c^4} \left( \frac{q_1}{|\mathbf{x}-\mathbf{x}_1|} + \frac{q_2}{|\mathbf{x}-\mathbf{x}_2|} \right)^2 + \frac{G}{c^4} \left( \frac{q_1 q_2}{r |\mathbf{x}-\mathbf{x}_1|} + \frac{q_1 q_2}{r |\mathbf{x}-\mathbf{x}_2|} \right),
        \end{split}
    \end{equation}
    \begin{equation}
        \begin{split}
            V_j &= \frac{G}{c^3} \left( \frac{m_1 v_1^j}{|\mathbf{x}-\mathbf{x}_1|} + \frac{m_2 v_2^j}{|\mathbf{x}-\mathbf{x}_2|} \right),
        \end{split}
    \end{equation}
    \begin{equation}\label{sca_nearzone}
   	\begin{split} 
   		\varphi &= -\frac{G}{c^2} \left( \frac{m_1 \alpha_1}{|\mathbf{x}-\mathbf{x}_1|} + \frac{m_2 \alpha_2}{|\mathbf{x}-\mathbf{x}_2|} \right) + \frac{G}{2c^4} m_1 \alpha_1 \left( \mathbf{n}_1 \cdot \mathbf{a}_1 + \frac{(\mathbf{n}_1 \cdot \mathbf{v}_1)^2}{|\mathbf{x}-\mathbf{x}_1|} \right) + \frac{G}{2c^4} m_2 \alpha_2 \left( \mathbf{n}_2 \cdot \mathbf{a}_2 + \frac{(\mathbf{n}_2 \cdot \mathbf{v}_2)^2}{|\mathbf{x}-\mathbf{x}_2|} \right) \\
   		&+ \frac{G a}{2c^4} \left( \frac{q_1}{|\mathbf{x}-\mathbf{x}_1|} + \frac{q_2}{|\mathbf{x}-\mathbf{x}_2|} \right)^2 - \frac{G a}{c^4} q_1 q_2 \left( \frac{1}{r |\mathbf{x}-\mathbf{x}_1|} + \frac{1}{r |\mathbf{x}-\mathbf{x}_2|} \right) \\
   		&+ \frac{G^2}{c^4} m_1 m_2 \left[ \frac{\alpha_1 + \alpha_2(\alpha_1^2 + \beta_1)}{r |\mathbf{x}-\mathbf{x}_1|} + \frac{\alpha_2 + \alpha_1(\alpha_2^2 + \beta_2)}{r |\mathbf{x}-\mathbf{x}_2|} \right],
   	\end{split}
   \end{equation}
    \begin{equation} \label{em0_nearzone}
        \begin{split}
            A_0 &= -\left(\frac{q_1}{\left|\mathbf{x}-\mathbf{x}_1\right|}+\frac{q_2}{\left|\mathbf{x}-\mathbf{x}_2\right|}\right) \\
            & -\frac{q_1}{2 c^2}\left(\frac{v_1^2}{\left|\mathbf{x}-\mathbf{x}_1\right|}-\mathbf{n}_1 \cdot \mathbf{a}_1-\frac{\left(\mathbf{n}_1 \cdot \mathbf{v}_1\right)^2}{\left|\mathbf{x}-\mathbf{x}_1\right|}\right)-\frac{q_2}{2 c^2}\left(\frac{v_2^2}{\left|\mathbf{x}-\mathbf{x}_2\right|}-\mathbf{n}_2 \cdot \mathbf{a}_2-\frac{\left(\mathbf{n}_2 \cdot \mathbf{v}_2\right)^2}{\left|\mathbf{x}-\mathbf{x}_2\right|}\right) \\
            & +\frac{G}{c^2}\left[\frac{\left(1+a \alpha_2\right) q_1 m_2}{r\left|\mathbf{x}-\mathbf{x}_1\right|}+\frac{\left(1+a \alpha_1\right) q_2 m_1}{r\left|\mathbf{x}-\mathbf{x}_2\right|}\right]-\frac{G}{c^2}\left[\frac{\left(1+a \alpha_1\right) m_1 q_2}{r\left|\mathbf{x}-\mathbf{x}_1\right|}+\frac{\left(1+a \alpha_2\right) m_2 q_1}{r\left|\mathbf{x}-\mathbf{x}_2\right|}\right] \\
            & +\frac{G}{c^2}\left(\frac{\left(1+a \alpha_1\right) m_1}{\left|\mathbf{x}-\mathbf{x}_1\right|}+\frac{\left(1+a \alpha_2\right) m_2}{\left|\mathbf{x}-\mathbf{x}_2\right|}\right)\left(\frac{q_1}{\left|\mathbf{x}-\mathbf{x}_1\right|}+\frac{q_2}{\left|\mathbf{x}-\mathbf{x}_2\right|}\right).
        \end{split}
    \end{equation}
    \begin{equation} \label{emj_nearzone}
        \begin{split}
            A_j &= \frac{1}{c} \left( \frac{q_1 v_1^j}{|\mathbf{x}-\mathbf{x}_1|} + \frac{q_2 v_2^j}{|\mathbf{x}-\mathbf{x}_2|} \right).
        \end{split}
    \end{equation}
\end{widetext}
Here $V$ and $V_j$ denote the scalar and vector post-Newtonian metric potentials entering the near-zone metric $g_{\mu \nu}$, originated from the Fokker action method. $m_A$, $\alpha_A$ ($\beta_A$) and $q_A$ are the mass, scalar charges and electric charge of particle $A$ (1 or 2 in this paper) respectively. For coordinate quantities, $\mathbf{x}$ denotes the position of the field point, $\mathbf{x}_A$ is the position of particle $A$, $\mathbf{v}_A$ and $\mathbf{a}_A$ are the corresponding velocity and acceleration of particle $A$. Finally, $\mathbf{n}_A=(\mathbf{x}-\mathbf{x}_A)/|\mathbf{x}-\mathbf{x}_A|$ denotes the unit direction vector pointing to the position of the field point $\mathbf{x}$ from the position of particle $\mathbf{x}_A$, and $r=|\mathbf{x}_1 - \mathbf{x}_2|$ is the relative distance between particle 1 and 2. Our DIRE calculation reproduces the near-zone fields obtained by Khalil et al. \cite{Khalil:2018aaj} using the Fokker-action method.

To evaluate the effective scalar source $\mu$ (cf. \eqref{Def_scalar_eff_source}) through relative 1PN order, only the leading $h^{00}$ component of the near-zone gothic metric perturbation is required. For later use in the electromagnetic and gravitational source moments, we also give $h^{0j}$. The spatial component  $h^{ij}$ starts at $\mc{O}(c^{-4})$ and contributes only beyond the accuracy needed here. The relevant components are 
\begin{widetext}
\begin{equation} \label{grav00_nearzone}
\begin{split}
    h^{00} &= \frac{4G}{c^2} \left( \frac{m_1}{|\mathbf{x}-\mathbf{x}_1|} + \frac{m_2}{|\mathbf{x}-\mathbf{x}_2|} \right) + \frac{2G}{c^4} \left( \frac{m_1 v_1^2}{|\mathbf{x}-\mathbf{x}_1|} + \frac{m_2 v_2^2}{|\mathbf{x}-\mathbf{x}_2|} \right) + \frac{2G}{c^4} m_1 \left( \frac{v_1^2}{|\mathbf{x}-\mathbf{x}_1|} - \mathbf{n}_1 \cdot \mathbf{a}_1 - \frac{(\mathbf{n}_1 \cdot \mathbf{v}_1)^2}{|\mathbf{x}-\mathbf{x}_1|} \right) \\
    &\quad + \frac{2G}{c^4} m_2 \left( \frac{v_2^2}{|\mathbf{x}-\mathbf{x}_2|} - \mathbf{n}_2 \cdot \mathbf{a}_2 - \frac{(\mathbf{n}_2 \cdot \mathbf{v}_2)^2}{|\mathbf{x}-\mathbf{x}_2|} \right) - \frac{G^2}{c^4} \left( \frac{m_1 \alpha_1}{|\mathbf{x}-\mathbf{x}_1|} + \frac{m_2 \alpha_2}{|\mathbf{x}-\mathbf{x}_2|} \right)^2 - \frac{2G^2}{c^4} \left( \frac{m_1 m_2 \alpha_1 \alpha_2}{r |\mathbf{x}-\mathbf{x}_1|} + \frac{m_1 m_2 \alpha_1 \alpha_2}{r |\mathbf{x}-\mathbf{x}_2|} \right) \\
    &\quad + \frac{7G^2}{c^4} \left( \frac{m_1}{|\mathbf{x}-\mathbf{x}_1|} + \frac{m_2}{|\mathbf{x}-\mathbf{x}_2|} \right)^2 - \frac{2G^2}{c^4} \left( \frac{m_1 m_2}{r |\mathbf{x}-\mathbf{x}_1|} + \frac{m_1 m_2}{r |\mathbf{x}-\mathbf{x}_2|} \right) - \frac{G}{c^4} \left( \frac{q_1}{|\mathbf{x}-\mathbf{x}_1|} + \frac{q_2}{|\mathbf{x}-\mathbf{x}_2|} \right)^2 \\
    &\quad + \frac{2G}{c^4} \left( \frac{q_1 q_2}{r |\mathbf{x}-\mathbf{x}_1|} + \frac{q_1 q_2}{r |\mathbf{x}-\mathbf{x}_2|} \right),
\end{split}
\end{equation}
\begin{equation} \label{grav0j_nearzone}
    h^{0j} = \frac{4G}{c^3} \left( \frac{m_1 v_1^j}{|\mathbf{x}-\mathbf{x}_1|} + \frac{m_2 v_2^j}{|\mathbf{x}-\mathbf{x}_2|} \right).
\end{equation}
\end{widetext}

\subsection{Scalar field}
We now construct the scalar field at a wave-zone field point generated by the effective scalar source inside the near zone, following the method of Refs.~\cite{Will:1996zj, Poisson_Will_2014}. Starting from the retarded solution (\ref{eq:scalar_formal_solution}), we restrict the source point \(\mathbf{x}'\) to the near-zone volume \(\mathcal{M}\), while the field point \(\mathbf{x}\) is taken to be far from the source, \(R=|\mathbf{x}|\gg \lambda_c\). In this regime, the Green function and the retarded time can be expanded in powers of \(1/R\). The resulting near-zone contribution to the wave-zone scalar field can then be written as a multipolar expansion in terms of the scalar source moments \(\Psi_L(\tau)\):
\begin{align} \label{sca_from_multipole_moment}
    \varphi &= \frac{G}{c^2} \sum_{\ell=0}^{\infty} \frac{1}{c^{\ell+2} \ell!} \left[ \frac{1}{R} N^L \Psi_L^{(\ell)} + \frac{1}{R^2} \right. \notag \\
    &\quad \left. \times \left( \frac{\ell+2}{2} N^{L+1} \Psi_{L+1}^{(\ell)} - \frac{\ell}{2} N^{L-1} \delta^{ab} \Psi_{ab(L-1)}^{(\ell)} \right) \right],
\end{align}
where $L=j_1\cdots j_\ell$ denotes a multi-index of length $\ell$, $\mathbf{N} = \mathbf{x}/R$ is the unit direction vector pointing to wave-zone field point, $N^L=N^{j_1}\cdots N^{j_\ell}$ is an abbreviation of collection of unit direction vectors, the superscript bracket is the $\ell$-th time derivative with respect to the retarded time $\tau=t-R/c$ and the scalar multipole moments $\Psi_L$ are defined as integrals of the effective scalar source $\mu$ over the near-zone volume $\mathcal{M}$
\begin{equation}
    \Psi_L(\tau) = \int_{\mathcal{M}} \mu(\tau, \mathbf{x}') x'^L d^3x'.
\end{equation}
In Eq.~\eqref{sca_from_multipole_moment}, the \(R^{-1}\) term gives the asymptotic radiative field, whereas the \(R^{-2}\) term represents the next-to-leading finite-distance correction in the wave-zone expansion.

By utilizing the known field quantities shown in Sec.~\ref{sec:sol_nearzone}, we can determine the specific components of the effective scalar source $\mu$ up to 1PN order
\begin{align}
    \mu &= - \rho_g c^2 \alpha \left[ 1 - \frac{v^2}{2c^2} - \frac{1}{4} h^{00} + \left( \alpha + \frac{\beta}{\alpha} \right) \varphi \right] \notag \\
    &\quad - \frac{a}{4\pi} (\nabla A_0)^2 + \mathcal{O}(c^{-2}),
\end{align}
where $\rho_g$ is the mass density defined in \eqref{rhogandrhoe}.

In the process of calculating the scalar multipole moments (as well as the gravitational tensor and electromagnetic vector moments), we utilize integration by parts (see Ref. \cite{Will:1996zj}, Eq. (4.1) or Ref. \cite{Shiralilou:2021mfl}, Eq. (22)) to transform any volume integrals of the form $\int_{\mathcal{M}} (\nabla \varphi)^2 d^3x$ into a combination of surface integrals and alternative volume integrals. These surface integrals involve the boundary radius $\mathcal{R}$. Since the demarcation between the near zone and the wave zone is an artificial construct, the physical results must be independent of $\mathcal{R}$ \cite{Will:1996zj}. This principle allows for the cancellation of all $\mathcal{R}$-dependent terms. For the scalar multipole moments required in our analysis, the result is given by
\begin{align}
    \Psi &= - \sum_{A} \bigg[ m_A c^2 \alpha_A \bigg( 1 - \frac{v_A^2}{2c^2} - \sum_{B \neq A} \frac{G m_B}{c^2 r_{AB}} \notag \\
    &\quad - \Big( \alpha_A + \frac{\beta_A}{\alpha_A} \Big) \sum_{B \neq A} \frac{G m_B \alpha_{B}}{c^2 r_{AB}} \bigg) + \sum_{B \neq A} \frac{a q_A q_B}{r_{AB}}\bigg] \notag \\
    &\quad + \mathcal{O}(c^{-2}), \label{Psi} \\
    \Psi_j &= - \sum_{A} \bigg[ m_A c^2 \alpha_A \bigg( 1 - \frac{v_A^2}{2c^2} - \sum_{B \neq A} \frac{G m_B}{c^2 r_{AB}} \notag \\
    &\quad - \Big( \alpha_A + \frac{\beta_A}{\alpha_A} \Big) \sum_{B \neq A} \frac{G m_B \alpha_{B}}{c^2 r_{AB}} \bigg) \notag \\
    &\quad + \sum_{B \neq A} \frac{a q_A q_B}{r_{AB}}\bigg] x_A^j + \mathcal{O}(c^{-2}), \label{Psij} \\
    \Psi_{jk} &= - \sum_{A} m_A c^2 \alpha_A x_A^j x_A^k + \mathcal{O}(1), \label{Psijk} \\
    \Psi_{jkl} &= - \sum_{A} m_A c^2 \alpha_A x_A^j x_A^k x_A^l + \mathcal{O}(1),\label{Psijkl}
\end{align}
where $m_A$, $v_A$, $q_A$, $\alpha_A$ and $\beta_A$ denote the mass, velocity, charge, and scalar charges of particle $A$, respectively, with $r_{AB} = |\mathbf{x}_A - \mathbf{x}_B|$. These scalar multipole moments incorporate the nonlinear field contributions.

\subsection{Electromagnetic field}

In the wave zone, following the method of Refs.~\cite{Will:1996zj, Poisson_Will_2014} (see also Ref.~\cite{Damour:1990gj} for an alternative formalism), the electromagnetic 4-potential $A^{\nu}$ is obtained by solving the relaxed electromagnetic equations using the retarded Green's function. In the wave zone, the electric potential $A^0$ is expanded in terms of electric multipole moments $Q_L(\tau)$ as
\begin{align}\label{wavezoneA0}
    A^0 &= \sum_{\ell=0}^{\infty} \frac{1}{c^{\ell} \ell!} \left[ \frac{1}{R} N^L Q_L^{(\ell)} + \frac{1}{R^2} \right. \notag \\
    &\quad \left. \times \left( \frac{\ell+2}{2} N^{L+1} Q_{L+1}^{(\ell)} - \frac{\ell}{2} N^{L-1} \delta^{ab} Q_{ab(L-1)}^{(\ell)} \right) \right],
\end{align}
where $\tau = t - R/c$ is the retarded time and $N^L$ denotes the multi-index unit direction vectors.

The spatial components of the 4-potential (magnetic vector potentials), $A^j$, include contributions from the magnetic multipole moments $M_L(\tau)$. Its wave-zone expansion is given by
\begin{align}\label{wavezoneAj}
    A^j &= \sum_{\ell=0}^{\infty} \frac{1}{c^{\ell} \ell!} \left[ \frac{1}{R} N^L M_{jL}^{(\ell)} + \frac{1}{R^2} \right. \notag \\
    &\quad \left. \times \left( \frac{\ell+2}{2} N^{L+1} M_{j(L+1)}^{(\ell)} - \frac{\ell}{2} N^{L-1} \delta^{ab} M_{jab(L-1)}^{(\ell)} \right) \right].
\end{align}

The source multipole moments are related to the effective source $\mu^0$ and $\mu^j$. The electric multipole moments $Q_L$ and magnetic multipole moments $M_L$ are defined as
\begin{align}\label{QL}
    Q_L(\tau) &=\int_{\mathcal{M}} \mu^0(\tau, \mathbf{x}') x'^L d^3x', \\
    M_{jL}(\tau) &= \int_{\mathcal{M}} \mu^j(\tau, \mathbf{x}') x'^L d^3x'.\label{MjL}
\end{align}

Following the approach established in the previous subsection, we can directly obtain the effective source $\mu^0$ and $\mu^j$ up to 1PN order
\begin{align}
    \mu^0 =& (1 + 2a \varphi) \rho_e + \frac{a}{2\pi} \partial_j \varphi \partial_j A_0 + \frac{1}{8\pi} \partial_j (A_0 \partial_j h^{00}) \\ \notag
    & + \mathcal{O}(c^{-4 }), \\
    \mu^j =& (1 + 2a \varphi) \frac{\rho_e v^j}{c} - \frac{a}{2\pi c} \dot{\varphi} \partial_j A_0 \notag \\
    &- \frac{a}{2\pi} \partial_k \varphi (\partial_k A_j - \partial_j A_k) \\ \notag
    & - \frac{1}{8\pi} \partial_0 (A_0 \partial_j h^{00}) + \frac{1}{4\pi}\partial_k[A_0 (\partial_k h^{0j} - \partial_j h^{0k})] \\ \notag
    & + \frac{1}{8\pi} \partial_k(A_j \partial_k h^{00} - A_k \partial_j h^{00})+ \mathcal{O}(c^{-5}),
\end{align}
where $\rho_e$ is the electric charge density defined in \eqref{rhogandrhoe}, $a$ is the dilaton coupling constant, and the second term of each equation accounts for the nonlinear interaction between the scalar field $\varphi$ and the electromagnetic field in the near zone.

By substituting these effective sources into the multipole integrals, we obtain the electric multipole moments $Q_L$. The monopole and dipole moments, which incorporate 1PN corrections, along with the higher-order moments at leading order, are expressed as follows:
    \begin{align}
        Q_{\text{mon}} &= \sum_A q_A + \mathcal{O}(c^{-4}), \\
        Q_j &= \sum_A q_A \left( x_A^j - \sum_{B \neq A} \frac{G m_B (1 + a \alpha_B) r_{AB}^j}{c^2 r_{AB}} \right) \\ \notag
        & + \mathcal{O}(c^{-4}), \\
        Q_{jk} &= \sum_A q_A x_A^j x_A^k + \mathcal{O}(c^{-2}), \\
        Q_{jkl} &= \sum_A q_A x_A^j x_A^k x_A^l + \mathcal{O}(c^{-2}),
    \end{align}
where $q_A$ is the constant electric charge of particle $A$.

Similarly, the magnetic multipole moments are obtained by integrating the effective magnetic source $\mu^j$. For the magnetic dipole and quadrupole moments, which include the 1PN corrections, and the higher-order moments at leading order, we have:
\begin{align}
    M_j &= \sum_A \frac{q_A v_A^j}{c} - \frac{G}{c^3} \sum_A \sum_{B \neq A} \frac{q_A m_B (1 + a \alpha_B)}{r_{AB}} \\ \notag
    & \times\left[ v_A^j - v_B^j  - \frac{(\mathbf{v}_A - \mathbf{v}_B) \cdot \mathbf{r}_{AB}}{r_{AB}^2} r_{AB}^j \right] + \mathcal{O}(c^{-5}), \\
    M_{jk} &= \sum_A \frac{q_A v_A^j x_A^k}{c}
- \frac{G}{c^3} \sum_A \sum_{B \neq A} q_A m_B (1+a\alpha_B) \notag \\
& \times \Bigg\{ v_A^j n_{AB}^k
+ x_A^k \frac{v_{AB}^j - (\mathbf{v}_{AB} \cdot \mathbf{n}_{AB}) n_{AB}^j}{r_{AB}} \notag \\
&\quad + \frac{1}{2} \Big[ \delta_{jk} (\mathbf{v}_{AB} \cdot \mathbf{n}_{AB})
- v_{AB}^k n_{AB}^j - v_{AB}^j n_{AB}^k \notag \\
&\qquad + (\mathbf{v}_{AB} \cdot \mathbf{n}_{AB}) n_{AB}^j n_{AB}^k \Big] \Bigg\} \notag \\
& - \frac{G}{c^3} \sum_A \sum_{B \neq A} q_A m_B
\Big[2\big(v_{AB}^k n_{AB}^j-v_{AB}^j n_{AB}^k\big) \\ \notag
&\quad -\delta_{jk}(\mathbf{v}_{AB}\cdot\mathbf{n}_{AB})\Big] + \mathcal{O}(c^{-5}), \\
    M_{jkl} &= \sum_A \frac{q_A v_A^j x_A^k x_A^l}{c} + \mathcal{O}(c^{-3}), \\
    M_{jklm} &= \sum_A \frac{q_A v_A^j x_A^k x_A^l x_A^m}{c} + \mathcal{O}(c^{-3}).
\end{align}

We emphasize that the moments $Q_L$ and $M_{jL}$ defined in
Eqs.~(\ref{QL}) and (\ref{MjL}) are direct Cartesian moments of the
effective electromagnetic source and are not, in general, irreducible
symmetric-trace-free (STF) electric- and magnetic-type multipole
moments. They therefore need not agree term by term with the
electromagnetic multipoles introduced in Ref.~\cite{Khalil:2018aaj}. The two
representations are nevertheless equivalent after the Cartesian
moments are decomposed into their STF parts and traces. In particular,
a trace of a Cartesian moment of rank $\ell$ contains lower-multipole
components. Consequently, when the flux is evaluated directly in the
present non-STF representation, cross terms between moments of
different tensor ranks, such as monopole--quadrupole terms, may appear
and give nonvanishing contributions. In an irreducible STF
representation, these lower-multipole pieces are separated explicitly,
and contributions with different values of $\ell$ are orthogonal under
the angular integration. After all trace contributions are properly
combined, the reconstructed electromagnetic field and the resulting
physical fluxes are independent of the choice of multipolar
representation \cite{Damour:1990gj,Blanchet:2013haa}.

\subsection{Gravitational field} \label{sec:grav_field}

In the wave zone, the spatial gravitational field $h^{jk}$ that contains the main physical information is obtained by solving the relaxed Einstein field equations with the retarded Green's function. To organize the wave-zone expansion, we employ the Epstein-Wagoner multipole formalism, which is well suited for treating the noncompact effective gravitational source $\mu^{jk}$ and for expressing the radiative field in terms of source moments~\cite{Will:1996zj}. At leading order in the inverse distance expansion, the standard Epstein-Wagoner moments provide the compact form of the $\mathcal{O}(R^{-1})$ waveform. For the next-to-leading $\mathcal{O}(R^{-2})$ terms, the same multipolar expansion naturally brings in higher multipole moments, beginning with $I^{jkl}$, together with their time derivatives. We therefore keep the Epstein-Wagoner organization of the retarded solution while retaining the higher-order source moments required at this order. Applying the shortwave approximation and expanding the retarded solution through $\mathcal{O}(R^{-2})$, we obtain

\begin{align}\label{GWwaveform}
    h^{jk} &= \frac{4G}{c^4} \sum_{\ell=0}^{\infty} \frac{1}{\ell! c^\ell} \left[ \frac{1}{R} N^L I^{jk(\ell)}_L + \frac{1}{R^2} \right. \notag \\
    &\quad \left. \times \left( \frac{\ell+2}{2} N^{L+1} I^{jk(\ell)}_{L+1} - \frac{\ell}{2} N^{L-1} \delta^{ab} I^{jk(\ell)}_{ab(L-1)} \right) \right],
\end{align}
where the gravitational multipole moments $I^{jk}_L$ are defined as integrals of the effective gravitational source $\mu^{jk}$ over the near-zone volume $\mathcal{M}$
\begin{equation}\label{IjkL}
    I^{jk}_L(\tau) = \int_{\mathcal{M}} \mu^{jk}(\tau, \mathbf{x}') x'^L d^3x'.
\end{equation}

Following the Epstein-Wagoner moment formalism, for the computation of $I^{jk}$ and $I^{jk}_l$, the effective gravitational source $\mu^{jk}$ can be directly represented by $\mu^{00}$ and $\mu^{0j}$, which are obtained using Eq.~\eqref{Def_gravitational_eff_source}:
\begin{align}
        \mu^{00} &= \rho_g c^2 \left( 1 + \frac{v^2}{2c^2} + \frac{3}{4}h^{00} + \alpha\varphi \right) + \frac{c^4}{8\pi G} \partial_i\varphi \partial_i\varphi \notag \\
        &\quad + \frac{1}{8\pi} \partial_i A_0 \partial_i A_0 - \frac{7c^4}{128\pi G} \partial_i h^{00} \partial_i h^{00} + \mathcal{O}(c^{-2}), \\
        \mu^{0j} &= \rho_g c v^j \left( 1 + \frac{v^2}{2c^2} + \frac{3}{4}h^{00} + \alpha\varphi \right) \notag \\
        &\quad - \frac{c^3}{4\pi G} \dot{\varphi} \partial_j \varphi + \frac{1}{4\pi} \partial_i A_0 (\partial_i A_j - \partial_j A_i) \notag \\
        &\quad + \frac{c^4}{16\pi G} \left[ \frac{3}{4} \partial_j h^{00} \partial_0 h^{00} + (\partial_j  h^{0k} - \partial_k h^{0j}) \partial_k h^{00} \right] \notag \\
        &\quad + \mathcal{O}(c^{-3}).
\end{align}

To compute the gravitational multipole moments $I^{jk}$ and $I^{jk}_l$, we employ the conservation laws from Ref.~\cite{Will:1996zj}, as presented in Eqs.~(2.17a) and (2.17b):
\begin{align}
    \mu^{jk} &= \frac{1}{2} (\mu^{00} x^j x^k)_{,00} + 2 (\mu^{l(j} x^{k)})_{,l} - \frac{1}{2} (\mu^{mn} x^j x^k)_{,mn}, \label{eq:EW1} \\
    \mu^{jk} x^l &= \frac{1}{2} (2 \mu^{0(j} x^{k)} x^l - \mu^{0l} x^j x^k)_{,0} \notag \\
    &\quad + \frac{1}{2} (2 \mu^{m(j} x^{k)} x^l - \mu^{lm} x^j x^k)_{,m}. \label{eq:EW2}
\end{align}
Here, a comma followed by spacetime indices denotes partial differentiation with respect to the corresponding coordinates. Parentheses around indices denote symmetrization with weight one.

For the mass quadrupole moment $I^{jk}$, the volume integral over the effective source component $\mu^{jk}$ is rewritten as
\begin{align}
    I^{jk} &= \frac{1}{2 c^2} \frac{d^2}{dt^2} \int_{\mathcal{M}} \mu^{00} x^j x^k d^3x \notag \\
    &\quad + \oint_{\partial\mathcal{M}} \left[ 2 \mu^{l(j} x^{k)} - \frac{1}{2} \partial_m (\mu^{ml} x^j x^k) \right] dS_l.
\end{align}
For the mass octupole moment $I^{jk}_l$, the relation yields
\begin{align}
    I^{jk}_l &= \frac{1}{2 c} \frac{d}{dt} \int_{\mathcal{M}} \left( \mu^{0j} x^k x^l + \mu^{0k} x^j x^l - \mu^{0l} x^j x^k \right) d^3x \notag \\
    &\quad + \frac{1}{2} \oint_{\partial\mathcal{M}} \left( 2 \mu^{m(j} x^{k)} x^l - \mu^{lm} x^j x^k \right) dS_m.
\end{align}

The surface integrals in the above expressions can be estimated by their \(\mathcal R\)-dependence. Up to the 1PN order required here, these terms contain only explicit powers of the artificial boundary radius \(\mathcal R\) and do not generate any \(\mathcal R\)-independent contribution. Since physical source moments must be independent of the near-zone boundary, these boundary-dependent pieces are discarded. Therefore, only the volume-integral parts need to be evaluated in the following calculation.

To evaluate the mass hexadecapole moment $I^{jk}_{lm}$ and the mass 32-pole moment $I^{jk}_{lmn}$, we integrate the effective gravitational source $\mu^{jk}$ directly. The spatial components of the effective gravitational source are given by
\begin{align}
    \mu^{jk} &= \rho_g v^j v^k + \frac{c^4}{4\pi G} \left[ \partial_j\varphi \partial_k\varphi - \frac{1}{2}\delta_{jk} (\partial_i\varphi \partial_i\varphi) \right] \notag \\
        &\quad + \frac{1}{4\pi} \left[ \frac{1}{2}\delta_{jk} (\partial_i A_0 \partial_i A_0) - \partial_j A_0 \partial_k A_0 \right] \notag \\
        &\quad + \frac{c^4}{64\pi G} \left( \partial_j h^{00}\partial_k h^{00} - \frac{1}{2}\delta_{jk} \partial_i h^{00}\partial_i h^{00} \right) \notag \\
        &\quad + \mathcal{O}(c^{-2}).
\end{align}

The leading-order and 1PN gravitational moments are evaluated as
\begin{align}
    I^{jk} &= \frac{1}{2} \frac{d^2}{dt^2} \Bigg\{ \sum_A m_A x_A^j x_A^k \bigg[ 1 + \frac{v_A^2}{2c^2} \notag \\
        &\quad - \sum_{B \neq A} \left( \frac{G m_B}{2c^2 r_{AB}} (1 + \alpha_A \alpha_B) - \frac{q_A q_B}{2 m_A c^2 r_{AB}} \right) \bigg] \notag \\
        &\quad + \frac{1}{4c^2} \delta_{jk} \sum_{B \neq A} \left[ G m_A m_B (7 - \alpha_A \alpha_B) - q_A q_B \right] r_{AB} \bigg\} \notag \\
        &\quad + \mathcal{O}(c^{-4}), \\
        I^{jk}_l &= I^{jk}_{l,V,m} + I^{jk}_{l,V,\varphi} + I^{jk}_{l,V,A} + I^{jk}_{l,V,\text{LL}} + \mathcal{O}(c^{-4}), \\
        I^{jk}_{lm} &= \sum_{A} m_{A} \bigg[ \sum_{B \ne A} G m_B \bigg( 1 + \alpha_A \alpha_B - \frac{q_A q_B}{G m_A m_B} \bigg) \notag \\
        &\quad \times  D_{jklm} (A,B) +v_{A}^{j} v_{A}^{k} x_{A}^{l} x_{A}^{m} \bigg] + \mathcal{O}(c^{-2}), \\
        I^{jk}_{lmn} &= \sum_{A} m_{A} \bigg[ \sum_{B \ne A} G m_B \bigg( 1 + \alpha_A \alpha_B - \frac{q_A q_B}{G m_A m_B} \bigg) \notag \\
        &\quad \times  E_{jklmn} (A,B) +v_{A}^{j} v_{A}^{k} x_{A}^{l} x_{A}^{m} x_{A}^{n} \bigg] + \mathcal{O}(c^{-2}),
\end{align}
where the mass octupole moment $I^{jk}_l$ receives contributions from the matter source $I^{jk}_{l,V,m}$, the scalar field $I^{jk}_{l,V,\varphi}$, the electromagnetic field $I^{jk}_{l,V,A}$ and the nonlinear gravitational field $I^{jk}_{l,V,\text{LL}}$ represented by the Landau-Lifshitz pseudotensor. The harmonic gauge term $t_H^{jk}$ does not contribute to the mass octupole moment, since its contribution exceeds the accuracy we consider in this paper. The expression for these contributions and the coefficients that depend on the particle labels, $D_{ajkl}(A,B)$ and $E_{ajklm}(A,B)$, are deferred to Appendix \ref{app:grav_moment_coeff} for compactness.

A similar remark applies to the gravitational moments used above. The
Epstein--Wagoner moments $I^{jk}_{L}$ defined in Eq.~(\ref{IjkL})
are direct Cartesian source moments of the effective gravitational
stress and should not be identified term by term with the irreducible
STF mass- and current-type moments conventionally denoted by $I_L$ and
$J_L$. The latter are obtained only after imposing the conservation
relations, decomposing the Cartesian tensors into their STF parts and
traces, and reorganizing the resulting waveform into irreducible
multipoles. Accordingly, individual contributions associated with
different tensor ranks in the Epstein--Wagoner representation need not
correspond one-to-one to individual mass or current multipoles in the
STF representation. Nevertheless, both constructions yield the same
transverse-traceless radiative field after the complete multipolar
decomposition is performed, and hence the same gravitational-wave
observables and fluxes \cite{Damour:1990gj,Blanchet:2013haa}.
\section{Multipole Moments and Waveforms in the Center of Mass Frame} \label{sec:CM_Waveforms}

In this section, we construct the 1PN dynamics for the binary BH system and reduce the multipole moments derived in Sec.~\ref{sec:formal_solution} to the two-body center of mass (CM) frame. This reduction is essential for evaluating the angular momentum flux. The waveform expressions constructed from the multipole moments in this section represent the contributions $f_\mathcal{M}$ generated by the effective sources within the near zone. The additional contributions $f_\mathcal{W}$ generated by wave-zone sources are evaluated separately in Appendix~\ref{app:wavezone_contribution}; since they start at order $R^{-2}$, they do not modify the radiative $R^{-1}$ fields entering the angular momentum flux.

\subsection{Center-of-mass frame and 1PN dynamics}

The CM frame of the binary system is defined by the condition that the total mass dipole moment of the source vanishes, namely
\begin{equation}
    \int \mu^{00} x'^j d^3x' = 0,
\end{equation}
where $\mu^{00}$ is the effective energy density of the relaxed field equations. For a binary system of point masses in EMd theory, this condition determines the 1PN positions of the two bodies $\mathbf{x}_1$ and $\mathbf{x}_2$ in terms of the relative separation vector $\mathbf{r} = \mathbf{x}_1 - \mathbf{x}_2$ as
\begin{align}
    \mathbf{x}_1 &= \left[ \frac{m_2}{M} + \frac{\nu (m_1 - m_2)}{2 M c^2} \left( v^2 - \frac{G_{12} M}{r} \right) \right] \mathbf{r} + \mathcal{O}(c^{-4}), \label{eq:x1} \\
    \mathbf{x}_2 &= \left[ -\frac{m_1}{M} + \frac{\nu (m_1 - m_2)}{2 M c^2} \left( v^2 - \frac{G_{12} M}{r} \right) \right] \mathbf{r} + \mathcal{O}(c^{-4}), \label{eq:x2}
\end{align}
where $M = m_1 + m_2$ is the total mass, $\eta = m_1 m_2 / M$ is the reduced mass, $\nu = \eta / M$ is the symmetric mass ratio, and $G_{12} = G(1 + \alpha_1 \alpha_2 - q_1 q_2/G M \eta)$ is the effective coupling constant. The relative velocity is given by $\mathbf{v} = \dot{\mathbf{r}}$.

To obtain the two-body system Lagrangian for constructing dynamics, one can substitute the field quantities (Eqs.~\eqref{sca_nearzone}-\eqref{grav0j_nearzone}) in the near zone region into the action defined by Eq.~\eqref{Def_action}. Another way is applying Fokker action method, i.e., eliminating the redundant degrees of freedom in the metric $g_{\mu \nu}$ at first, then substituting the simplified field quantities into the Fokker action, obtaining the same two-body system Lagrangian (see Refs. \cite{Khalil:2018aaj, Julie:2017rpw}). The two-body system Lagrangian is
\begin{multline}
L=-m_1c^2-m_2c^2+\frac12m_1v_1^2+\frac12m_2v_2^2 \\
{}+\frac{Gm_1m_2}{r}(1+\alpha_1\alpha_2)-\frac{q_1q_2}{r} \\
{}+\frac{1}{c^2}\Bigg\{\frac18m_1v_1^4+\frac18m_2v_2^4 \\
{}+\frac{q_1q_2}{2r}\Big[\mathbf v_1\!\cdot\!\mathbf v_2
 +(\mathbf n\!\cdot\!\mathbf v_1)(\mathbf n\!\cdot\!\mathbf v_2)\Big] \\
{}+\frac{Gm_1m_2}{2r}\Big[(3-\alpha_1\alpha_2)(v_1^2+v_2^2) \\
{}-(7-\alpha_1\alpha_2)(\mathbf v_1\!\cdot\!\mathbf v_2) \\
{}-(1+\alpha_1\alpha_2)
 (\mathbf n\!\cdot\!\mathbf v_1)(\mathbf n\!\cdot\!\mathbf v_2)\Big] \\
{}-\frac{G^2m_1m_2}{2r^2}\Big[(1+2\alpha_1\alpha_2)(m_1+m_2) \\
{}+m_1\alpha_1^2(\alpha_2^2+\beta_2)
 +m_2\alpha_2^2(\alpha_1^2+\beta_1)\Big] \\
{}+\frac{Gq_1q_2}{r^2}\Big[m_1(1+a\alpha_1)+m_2(1+a\alpha_2)\Big] \\
{}-\frac{G}{2r^2}\Big[m_1q_2^2(1+a\alpha_1)
 +m_2q_1^2(1+a\alpha_2)\Big]\Bigg\}
 +\mathcal O(c^{-4}).
\end{multline}

To derive the effective one-body Lagrangian $L$ in the CM frame, we substitute the CM coordinate relations Eqs.~\eqref{eq:x1} and \eqref{eq:x2} into the two-body Lagrangian $L = -m_1 c^2 - m_2 c^2 + L_0 + c^{-2} L_1$. A crucial simplification arises because the 1PN coordinate correction only needs to be applied to the Newtonian part $L_0$, while the 0PN relations suffice for $L_1$.

The Newtonian Lagrangian $L_0$ is given by
\begin{equation}
    L_0 = \frac{1}{2} m_1 v_1^2 + \frac{1}{2} m_2 v_2^2 + \frac{G_{12} m_1 m_2}{r},
\end{equation}
where $G_{12}$ is the effective gravitational constant.

Substituting Eqs. (\ref{eq:x1}) and (\ref{eq:x2}) into the Newtonian Lagrangian $L_0$, the common 1PN shift in $\mb{x}_1$ and $\mb{x}_2$ cancels in the relative separation $\mb{r}=\mb{x}_1-\mb{x}_2$. In the kinetic term, the terms linear in the 1PN velocity correction cancel, while the quadratic terms are of order $\mc{O}(c^{-4})$ and can be neglected at 1PN order. Therefore $L_0$ elegantly reduces to the  Newtonian CM form
\begin{equation}
    L_{0,\text{CM}} = \frac{1}{2} \eta v^2 + \frac{G_{12} M \eta}{r}.
\end{equation}

For the 1PN Lagrangian $L_1$, we simply substitute the 0PN relations $\mathbf{v}_1 = m_2\mathbf{v}/M$ and $\mathbf{v}_2 = -m_1\mathbf{v}/M$. The effective one-body Lagrangian $L$ up to 1PN order becomes
\begin{multline}
L=-Mc^2+\frac12\eta v^2+\frac{G_{12}M\eta}{r}
 +\frac{1-3\nu}{8c^2}\eta v^4 \\
{}+\frac{G_{12}M\eta}{2c^2r}
 \left(\Gamma v^2+\nu\dot r^2\right)
 -\frac{G^2M^2\eta}{2c^2r^2}\Pi
 +\mathcal O(c^{-4}).
\end{multline}
where the total mass $M$, the reduced mass $\eta$ and the symmetric mass ratio $\nu$ have been defined previously. Besides $X_1 = m_1 / M$ and $X_2 = m_2 / M$ are mass ratios, and two parameters $\Gamma$ and $\Pi$ are defined by
\begin{align}
    \Gamma &= \frac{3 - \alpha_1 \alpha_2}{1 + \alpha_1 \alpha_2 - q_1 q_2 / G M \eta} + \nu, \\
    \Pi &= (1 + \alpha_1 \alpha_2)^2 + X_1 \alpha_1^2 \beta_2 + X_2 \alpha_2^2 \beta_1 \notag \\
    &\quad - 2 \frac{q_1 q_2}{G M \eta} (1 + a \alpha_1 X_1 + a \alpha_2 X_2) \notag \\
    &\quad + X_1 \frac{q_2^2}{G M \eta} (1 + a \alpha_1) + X_2 \frac{q_1^2}{G M \eta} (1 + a \alpha_2).
\end{align}

We substitute the effective one-body Lagrangian into the Euler–Lagrange equations and obtain the 1PN effective equation of motion
\begin{multline}\label{eq:EoM}
\mathbf a=-\frac{G_{12}M}{r^2}\mathbf n
 +\frac{G_{12}M}{c^2r^2}\Bigg\{
 \dot r(1-3\nu+\Gamma)\mathbf v \\
{}+\Bigg[\left(\frac12-\frac52\nu-\frac12\Gamma\right)v^2 \\
{}+\left(\Gamma+\nu+\frac{G^2}{G_{12}^2}\Pi\right)
 \frac{G_{12}M}{r}
 +\frac32\nu\dot r^2\Bigg]\mathbf n\Bigg\}
 +\mathcal O(c^{-4}).
\end{multline}
where $\mathbf{a} = d^2 \mathbf{r} / dt^2$ is the acceleration of effective one-body, $\mathbf{n} = \mathbf{r} / r$ is the unit direction vector of effective one-body's position. 

\subsection{Scalar waveform}

The wave-zone scalar field begins with a monopole moment $\Psi$. Unlike quasicircular orbits where $\dot{r}=0$, for general eccentric orbits, the radial motion induces a nonvanishing monopole energy flux \cite{Khalil:2018aaj}, however, the monopole angular momentum flux still does not exist \cite{Zhang:2018prg, Jain:2024lie}. Substituting the 1PN CM relations into the scalar multipole moments (Eqs.~\eqref{Psi}-\eqref{Psijkl}), we obtain
\begin{align}
    \Psi &= - M (X_1 \alpha_1 + X_2 \alpha_2) c^2 + \frac{1}{2} \eta (X_2 \alpha_1 + X_1 \alpha_2) v^2 \notag \\
    &\quad + \frac{G M \eta}{r} \bigg[ \alpha_1 + \alpha_2 + (\alpha_1^2 + \beta_1) \alpha_2 + (\alpha_2^2 + \beta_2) \alpha_1 \notag \\
    &\quad - \frac{2 a q_1 q_2}{G M \eta} \bigg] + \mathcal{O}(c^{-2}), \label{CM_Psi} \\
    \Psi_j &= - \eta (\alpha_1 - \alpha_2) c^2 r^j - M (X_1 \alpha_1 + X_2 \alpha_2) \delta r^j \notag \\
    &\quad + \frac{\eta}{2} (X_2^2 \alpha_1 - X_1^2 \alpha_2) v^2 r^j \notag \\
    &\quad + \frac{G M \eta}{r} \bigg[ X_2 \Big( \alpha_1 + \alpha_1^2 \alpha_2 + \beta_1 \alpha_2 - \frac{a q_1 q_2}{G M \eta} \Big) \notag \\
    &\quad - X_1 \Big( \alpha_2 + \alpha_2^2 \alpha_1 + \beta_2 \alpha_1 - \frac{a q_1 q_2}{G M \eta} \Big) \bigg] r^j + \mathcal{O}(c^{-2}), \label{CM_Psij} \\
    \Psi_{jk} &= - \eta (X_2 \alpha_1 + X_1 \alpha_2) c^2 r^j r^k + \mathcal{O}(1), \label{CM_Psijk} \\
    \Psi_{jkl} &= - \eta (X_2^2 \alpha_1 - X_1^2 \alpha_2) c^2 r^j r^k r^l + \mathcal{O}(1), \label{CM_Psijkl}
\end{align}
where
\begin{align}
    \delta &= \frac{\nu (X_1 - X_2)}{2} \left( v^2 - \frac{G_{12} M}{r} \right)
\end{align}
is the 1PN correction for CM coordinates relations.

The moments $\Psi$, $\Psi_j$, $\Psi_{jk}$, and $\Psi_{jkl}$ are sufficient for the scalar field at the accuracy needed here. In the $R^{-1}$ part of Eq. (\ref{sca_from_multipole_moment}), moments of rank $\ell$ enter with an additional factor $c^{-\ell}$; therefore moments up to $\ell=3$ are required to reach 1.5PN order. In the $R^{-2}$ part, the same expansion involves moments of rank $\ell+1$, so the same set of moments is sufficient to obtain the finite-distance terms through 1PN order. Higher-rank moments, or higher-PN corrections to the moments listed above, contribute only beyond the accuracy considered in this work.

Substituting these scalar multipole moments into Eq.~\eqref{sca_from_multipole_moment}, we obtain straightforwardly
\begin{align}
    \varphi(t,\mathbf{x}) &= \frac{G}{c^2 R} [\varphi_{R^{-1},0} + c^{-1} \varphi_{R^{-1},0.5} + c^{-2} \varphi_{R^{-1},1} \notag \\
    &\quad + c^{-3} \varphi_{R^{-1},1.5}] \notag \\
    &\quad + \frac{G}{c^2 R^2} [\varphi_{R^{-2},0} + c^{-1} \varphi_{R^{-2},0.5} + c^{-2} \varphi_{R^{-2},1}] \notag \\
    &\quad + \mathcal{O}(c^{-6}),
\end{align}
where $\varphi_{R^{-1},0}$, $\varphi_{R^{-1},0.5}$, $\varphi_{R^{-1},1}$ and $\varphi_{R^{-1},1.5}$ denote leading order $\mathcal{O}(R^{-1})$ contributions up to $\mathcal{O}(c^{-3})$. 
For the angular momentum flux at null infinity, only the radiative $\mathcal{O}(R^{-1})$ part is required. We nevertheless display the $\mathcal{O}(R^{-2})$ terms through 1PN order, since they represent finite-distance corrections to the wave-zone scalar field and may be useful for future applications. The $\mathcal{O}(R^{-1})$ part is kept through 1.5PN order because the scalar angular momentum flux is quadratic in the radiative field and its relative 1PN correction receives contributions from the interference between the leading dipole waveform and its 1PN correction. Then, $\varphi_{R^{-2},0}$, $\varphi_{R^{-2},0.5}$ and $\varphi_{R^{-2},1}$ up to 1PN accuracy $\mathcal{O}(c^{-2})$ are next-to-leading order $\mathcal{O}(R^{-2})$ contributions to the waveforms. The contributions up to 1.5PN and leading $\mathcal{O}(R^{-1})$ order in the CM system can be written as
\begin{align}
\varphi_{R^{-1},0}
 &= -M(X_1\alpha_1+X_2\alpha_2), \\
\varphi_{R^{-1},0.5}
 &= -\eta(\alpha_1-\alpha_2)(\mathbf N\cdot\mathbf v), \\
\varphi_{R^{-1},1}
 &= \tfrac12\eta(X_2\alpha_1+X_1\alpha_2)v^2 \notag\\
 &\quad+\frac{GM\eta}{r}\Big[\alpha_1+\alpha_2
 +(\alpha_1^2+\beta_1)\alpha_2 \notag\\
 &\qquad+(\alpha_2^2+\beta_2)\alpha_1
 -\frac{2a q_1q_2}{GM\eta}\Big] \notag\\
 &\quad-\eta(X_2\alpha_1+X_1\alpha_2)
 \Big[(\mathbf N\cdot\mathbf v)^2 \notag\\
 &\qquad-\frac{G_{12}M}{r^3}(\mathbf N\cdot\mathbf r)^2\Big], \\
\varphi_{R^{-1},1.5}
 &= -M(X_1\alpha_1+X_2\alpha_2)
 \Big[\dot\delta(\mathbf N\cdot\mathbf r) \notag\\
 &\qquad+\delta(\mathbf N\cdot\mathbf v)\Big] \notag\\
 &\quad+\tfrac12\eta(X_2^2\alpha_1-X_1^2\alpha_2)
 \Big[-\frac{2G_{12}M\dot r}{r^2}(\mathbf N\cdot\mathbf r) \notag\\
 &\qquad+v^2(\mathbf N\cdot\mathbf v)\Big] \notag\\
 &\quad+\frac{GM\eta}{r}\Bigg[
 X_2\left(\alpha_1+\alpha_1^2\alpha_2+\beta_1\alpha_2
 -\frac{a q_1q_2}{GM\eta}\right) \notag\\
 &\qquad-X_1\left(\alpha_2+\alpha_2^2\alpha_1+\beta_2\alpha_1
 -\frac{a q_1q_2}{GM\eta}\right)\Bigg] \notag\\
 &\qquad\times\left[-\frac{\dot r}{r}(\mathbf N\cdot\mathbf r)
 +(\mathbf N\cdot\mathbf v)\right] \notag\\
 &\quad-\eta(X_2^2\alpha_1-X_1^2\alpha_2)
 \Bigg[(\mathbf N\cdot\mathbf v)^3 \notag\\
 &\qquad-\frac{7G_{12}M}{2r^3}(\mathbf N\cdot\mathbf r)^2
 (\mathbf N\cdot\mathbf v) \notag\\
 &\qquad+\frac{3G_{12}M\dot r}{2r^4}
 (\mathbf N\cdot\mathbf r)^3\Bigg].
\end{align}
where the contribution at the 0.5PN order generates the angular-momentum dipole radiation, when the scalar charges $\alpha_1$ and $\alpha_2$ are equal, i.e., satisfying the dipole suppression limit, the leading scalar dipole contribution vanishes. The contributions up to 1PN and next-to-leading $\mathcal{O}(R^{-2})$ order in the CM system are
\begin{align}
\varphi_{R^{-2},0}
 &= -\eta(\alpha_1-\alpha_2)(\mathbf N\cdot\mathbf r), \\
\varphi_{R^{-2},0.5}
 &= -3\eta(X_2\alpha_1+X_1\alpha_2)
 (\mathbf N\cdot\mathbf v)(\mathbf N\cdot\mathbf r) \notag\\
 &\quad+\eta(X_2\alpha_1+X_1\alpha_2)r\dot r, \\
\varphi_{R^{-2},1}
 &= -M(X_1\alpha_1+X_2\alpha_2)\delta
 (\mathbf N\cdot\mathbf r) \notag\\
 &\quad+\frac{GM\eta X_2}{r} \notag\\
 &\quad\times\left(\alpha_1+\alpha_1^2\alpha_2+\beta_1\alpha_2
 -\frac{a q_1q_2}{GM\eta}\right)
 (\mathbf N\cdot\mathbf r) \notag\\
 &\quad-\frac{GM\eta X_1}{r} \notag\\
 &\quad\times\left(\alpha_2+\alpha_2^2\alpha_1+\beta_2\alpha_1
 -\frac{a q_1q_2}{GM\eta}\right)
 (\mathbf N\cdot\mathbf r) \notag\\
 &\quad+\eta(\alpha_2X_1^2-\alpha_1X_2^2)
 \Bigg[6(\mathbf N\cdot\mathbf v)^2(\mathbf N\cdot\mathbf r) \notag\\
 &\qquad-\frac{3G_{12}M}{r^3}(\mathbf N\cdot\mathbf r)^3
 -2r\dot r(\mathbf N\cdot\mathbf v) \notag\\
 &\qquad-\frac32v^2(\mathbf N\cdot\mathbf r)
 +\frac{3G_{12}M}{2r}(\mathbf N\cdot\mathbf r)\Bigg].
\end{align}

\subsection{Electromagnetic waveform}

The electromagnetic four-potential $A^\mu$ is similarly evaluated in the CM frame. The electric multipole moments $Q_L$ up to 1PN order take the form
\begin{align}
    Q_{\text{mon}} &= Q + \mathcal{O}(c^{-4}), \\
    Q_j &= (X_2 q_1 - X_1 q_2) r^j + Q \frac{\delta}{c^2} r^j \notag \\
    &\quad- \frac{G M}{c^2 r} [X_2 (1 + a \alpha_2) q_1 - X_1 (1 + a \alpha_1) q_2] r^j \\ \notag
    &\quad + \mathcal{O}(c^{-4}), \\
    Q_{jk} &= (X_2^2 q_1 + X_1^2 q_2) r^j r^k + \mathcal{O}(c^{-2}), \\
    Q_{jkl} &= (X_2^3 q_1 - X_1^3 q_2) r^j r^k r^l + \mathcal{O}(c^{-2}),
    \end{align}
while the magnetic multipole moments $M_{jL}$ up to 1PN order read

\begin{align}
    M_j &= (X_2 q_1 - X_1 q_2) \frac{v^j}{c} + Q \frac{\delta v^j + \dot{\delta} r^j}{c^3} \notag \\
    &\quad- \frac{G M}{c^3 r} \left[X_2 (1+a\alpha_2) q_1 - X_1 (1+a\alpha_1) q_2\right] \\ \notag
    &\qquad \times \bigg( v^j - \frac{\dot{r}}{r} r^j \bigg) + \mathcal{O}(c^{-5}), \\
    M_{jk} &= (X_2^2 q_1 + X_1^2 q_2) \frac{v^j}{c} r^k \notag \\
   &\quad + (X_2 q_1 - X_1 q_2) \frac{2 \delta v^j + \dot{\delta} r^j}{c^3} r^k \notag \\
   &\quad - \frac{G M}{c^3} \bigg[ 
   \big(X_2^2 (1+a\alpha_2) q_1 + X_1^2 (1+a\alpha_1) q_2\big) \\ \notag
   &\quad \times (2 v^j n^k - \dot{r} n^j n^k) \notag \\
   &\quad + \frac{1}{2} 
   \big(X_2 (1+a\alpha_2) q_1 + X_1 (1+a\alpha_1) q_2\big) \notag \\
   &\quad \times (\delta^{jk} \dot{r} - v^j n^k - n^j v^k + \dot{r} n^j n^k) \bigg] \notag \\
   &\quad - \frac{G M}{c^3} (X_2 q_1 + X_1 q_2)
   \Big[2(v^k n^j-v^j n^k)-\delta^{jk}\dot{r}\Big] \\ \notag
   &\quad + \mathcal{O}(c^{-5}), \\
    M_{jkl} &= (X_2^3 q_1 - X_1^3 q_2) \frac{v^j}{c} r^k r^l + \mathcal{O}(c^{-3}), \\
    M_{jklm} &= (X_2^4 q_1 + X_1^4 q_2) \frac{v^j}{c} r^k r^l r^m + \mathcal{O}(c^{-3}),
    \end{align}
and
\begin{equation}
    \dot{\delta} = - \frac{\nu G_{12} M (X_1 - X_2)}{2 r^2} \dot{r}.
\end{equation}
Here, $Q = q_1 + q_2$ is the total charge of two-body system, and we adopt notation $Q_{\text{mon}}$ to distinguish the electric monopole moment from the total electric charge $Q$. Similar to the PN counting demonstration in the previous subsection, the radiative $\mathcal{O}(R^{-1})$ part up to 1.5PN order needs moments whose rank up to $\ell=3$, and rank up to $l=3$ is sufficient for the next-to-leading part $\mathcal{O}(R^{-2})$ up to 1PN order. 

By substituting these multipole moments into the multipole expansion for the electromagnetic field in the wave zone Eqs. \eqref{wavezoneA0} and \eqref{wavezoneAj}, the components of the electromagnetic potential $A^0$ and $A^j$ are expressed as:
\begin{align}
    A^0 &= \frac{1}{R} [A^0_{R^{-1},0} + c^{-1} A^0_{R^{-1},0.5} + c^{-2} A^0_{R^{-1},1} \notag \\
    &\quad + c^{-3} A^0_{R^{-1},1.5}] \notag \\
    &\quad + \frac{1}{R^2} [A^0_{R^{-2},0} + c^{-1} A^0_{R^{-2},0.5} + c^{-2} A^0_{R^{-2},1}] \notag \\
    &\quad + \mathcal{O}(c^{-3}),
\end{align}

\begin{align}
    A^j &= \frac{1}{c R} [A_{R^{-1},0}^j + c^{-1} A_{R^{-1},0.5}^j + c^{-2} A_{R^{-1},1}^j \notag \\
    &\quad + c^{-3} A_{R^{-1},1.5}^j] \notag \\
    &\quad + \frac{1}{c R^2} [A_{R^{-2},0}^j + c^{-1} A_{R^{-2},0.5}^j + c^{-2} A_{R^{-2},1}^j] \notag \\
    &\quad + \mathcal{O}(c^{-4}).
\end{align}

The radiative $\mc O(R^{-1})$ electric potential required for the
electromagnetic angular-momentum flux is needed through $1.5$PN order.
A $1.5$PN electromagnetic scalar potential in Einstein--Maxwell theory
was recently obtained in Ref.~\cite{Placidi:2025xyi} in the context of
radiation-reaction dynamics. We stress that the potential considered
there is a near-zone PN quantity evaluated at the particle positions,
whereas $A^0_{R^{-1},1.5}$ below is the $\mc O(R^{-1})$ wave-zone
coefficient entering the radiative flux, so that the two expressions
are not directly comparable term by term. Its individual PN
contributions are
\begin{align}
A^0_{R^{-1},0} &= Q, \\
A^0_{R^{-1},0.5}
 &= (X_2q_1-X_1q_2)(\mathbf N\cdot\mathbf v), \\
A^0_{R^{-1},1}
 &= (X_2^2q_1+X_1^2q_2)
 \Big[(\mathbf N\cdot\mathbf v)^2 \notag\\
 &\qquad-\frac{G_{12}M}{r^3}(\mathbf N\cdot\mathbf r)^2\Big], \\
\end{align}
\begin{widetext}
\begin{align}
    A^0_{R^{-1},1.5}
 &= Q\Big[\delta(\mathbf N\cdot\mathbf v)
 +\dot\delta(\mathbf N\cdot\mathbf r)\Big] +\frac{GM}{r}\Big[(1+a\alpha_1)q_2X_1-(1+a\alpha_2)q_1X_2\Big]
 (\mathbf N\cdot\mathbf v) \\ \notag
 &\quad-\frac{GM\dot r}{r^2}
 \Big[(1+a\alpha_1)q_2X_1-(1+a\alpha_2)q_1X_2\Big](\mathbf N\cdot\mathbf r) \notag\\
 &\quad+(q_2X_1^3-q_1X_2^3)
 \Big[-(\mathbf N\cdot\mathbf v)^3 +\frac{7G_{12}M}{2r^3}(\mathbf N\cdot\mathbf r)^2
 (\mathbf N\cdot\mathbf v) -\frac{3G_{12}M\dot r}{2r^4}
 (\mathbf N\cdot\mathbf r)^3\Big].
\end{align}
and the 1PN contributions at the next-to-leading order are given by
\begin{equation}
    \begin{split}
        A^0_{R^{-2},0} &= (X_2 q_1 - X_1 q_2) (\mathbf{N} \cdot \mathbf{r}), \\
    A^0_{R^{-2},0.5} &= 3 (X_2^2 q_1 + X_1^2 q_2) (\mathbf{N} \cdot \mathbf{v}) (\mathbf{N} \cdot \mathbf{r}) - (X_2^2 q_1 + X_1^2 q_2) r \dot{r}, \\
    A^0_{R^{-2},1} &= Q\delta\,(\mathbf N\cdot\mathbf r)
+
\frac{GM}{r}
\left[
(1+a\alpha_1)q_2X_1-(1+a\alpha_2)q_1X_2
\right]
(\mathbf N\cdot\mathbf r)
\notag
\\
&\quad
+
\left(
q_2X_1^3-q_1X_2^3
\right)
\bigg[
\frac{3G_{12}M}{r^3}
(\mathbf N\cdot\mathbf r)^3
-6(\mathbf N\cdot\mathbf r)(\mathbf N\cdot\mathbf v)^2
+2r\dot r(\mathbf N\cdot\mathbf v)
+v^2(\mathbf N\cdot\mathbf r)
-\frac{3G_{12}M}{2r}
(\mathbf N\cdot\mathbf r)
\bigg],
    \end{split}
    \end{equation}
\end{widetext}
For the magnetic vector potential, we require 1PN contributions at leading order $\mathcal{O}(R^{-1})$ only to evaluate the angular momentum flux. However, we show the 1.5PN contributions here as references for further computation work:
\begin{widetext}
    \begin{align}
        A_{R^{-1},0}^j &= (X_2 q_1 - X_1 q_2) v^j, \\
        A_{R^{-1},0.5}^j &= (X_2^2 q_1 + X_1^2 q_2) \bigg[ (\mathbf{N} \cdot \mathbf{v}) v^j - \frac{G_{12} M}{r^2} (\mathbf{N} \cdot \mathbf{r}) n^j \bigg], \\
        A_{R^{-1},1}^j &=
v^j
\Bigg[
\delta Q
+\frac{GM}{r}
\Big((1+a\alpha_1)q_2X_1-(1+a\alpha_2)q_1X_2\Big) +
(q_2X_1^3-q_1X_2^3)
\bigg(
-(\mathbf N\cdot\mathbf v)^2
+\frac{3}{2}\frac{G_{12}M}{r^3}
(\mathbf N\cdot\mathbf r)^2
\bigg)
\Bigg]
\notag
\\
&\quad
+r^j
\Bigg[
\dot{\delta}Q
-\frac{GM\dot r}{r^2}
\Big((1+a\alpha_1)q_2X_1-(1+a\alpha_2)q_1X_2\Big) \\ \notag
&\qquad
+
G_{12}M(q_2X_1^3-q_1X_2^3)
\bigg(
\frac{2}{r^3}
(\mathbf N\cdot\mathbf r)(\mathbf N\cdot\mathbf v)
-\frac{3}{2}\frac{\dot r}{r^4}
(\mathbf N\cdot\mathbf r)^2
\bigg)
\Bigg], \\
        A^j_{R^{-1},1.5}
&=
-\frac{GM}{2r^2}
\left[
q_2X_1\left(1-a\alpha_1\right)
+
q_1X_2\left(1-a\alpha_2\right)
\right]
\left[
G_{12}M+r(\dot r^2-v^2)
\right]N^j
\notag \\
&\quad
+
\Bigg\{
-\left(q_2X_1-q_1X_2\right)
\left[
2\delta(\mathbf N\cdot\mathbf v)
+
3\dot\delta(\mathbf N\cdot\mathbf r)
\right]
\notag \\
&\qquad
+GM
\left[
\frac{\dot r}{r^2}
\left[
a\alpha_1q_2X_1(3X_1-1)
+
a\alpha_2q_1X_2(3X_2-1)
-
3\nu(q_1+q_2)
\right]
(\mathbf N\cdot\mathbf r)
\right.
\notag \\
&\qquad\qquad\left.
-\frac{1}{r}
\left[
(1+a\alpha_1)q_2X_1(X_1-X_2)
+
(1+a\alpha_2)q_1X_2(X_2-X_1)
\right]
(\mathbf N\cdot\mathbf v)
\right]
\notag \\
&\qquad
+\frac{MG_{12}\dot r}{r^2}
\left(
1-3\nu+\Gamma
\right)
\left(
q_2X_1^2+q_1X_2^2
\right)
(\mathbf N\cdot\mathbf r)
\notag \\
&\qquad
+
\left(
q_2X_1^4+q_1X_2^4
\right)
\left[
(\mathbf N\cdot\mathbf v)^3
-5\frac{G_{12}M}{r^3}
(\mathbf N\cdot\mathbf r)^2(\mathbf N\cdot\mathbf v)
+\frac{5}{2}\frac{G_{12}M\dot r}{r^4}
(\mathbf N\cdot\mathbf r)^3
\right]
\Bigg\}v^j
\notag \\
&\quad
+
\Bigg\{
\left(q_2X_1-q_1X_2\right)
\left[
2\frac{G_{12}M}{r^3}
\delta(\mathbf N\cdot\mathbf r)
-\dot\delta(\mathbf N\cdot\mathbf v)
-\ddot\delta(\mathbf N\cdot\mathbf r)
\right]
\notag \\
&\qquad
+GM
\left[
\frac{\dot r}{r^2}
\left[
q_2X_1
\left(
1+X_1-a\alpha_1X_2
\right)
+
q_1X_2
\left(
1+X_2-a\alpha_2X_1
\right)
\right]
(\mathbf N\cdot\mathbf v)
\right.
\notag \\
&\qquad\qquad\left.
+
\frac{1}{2r^4}
\left[
(1+a\alpha_1)q_2X_1(X_1-X_2)
+
(1+a\alpha_2)q_1X_2(X_2-X_1)
\right]
\left[
G_{12}M+r(-3\dot r^2+v^2)
\right]
(\mathbf N\cdot\mathbf r)
\right]
\notag \\
&\qquad
+\frac{M}{2r^4}
\left(
q_2X_1^2+q_1X_2^2
\right)
(\mathbf N\cdot\mathbf r)
\Big[
2G_{12}^2M(\nu+\Gamma)
+
G_{12}r
\left(
3\nu\dot r^2
+v^2
-5\nu v^2
-\Gamma v^2
\right)
+
2G^2M\Pi
\Big]
\notag \\
&\qquad
+
\left(
q_2X_1^4+q_1X_2^4
\right)
\left[
\frac{7}{6}\frac{G_{12}^2M^2}{r^6}
(\mathbf N\cdot\mathbf r)^3
+
\frac{1}{2}\frac{G_{12}M}{r^5}
(v^2-5\dot r^2)
(\mathbf N\cdot\mathbf r)^3
\right.
\notag \\
&\qquad\qquad\left.
+
\frac{9}{2}\frac{G_{12}M\dot r}{r^4}
(\mathbf N\cdot\mathbf r)^2(\mathbf N\cdot\mathbf v)
-
3\frac{G_{12}M}{r^3}
(\mathbf N\cdot\mathbf r)(\mathbf N\cdot\mathbf v)^2
\right]
\Bigg\}r^j .
\end{align}
\end{widetext}
and the 1PN contributions at the next-to-leading order are given by
\begin{widetext}
\begin{align}
    A_{R^{-2},0}^j &= (X_2^2 q_1 + X_1^2 q_2) (\mathbf{N} \cdot \mathbf{r}) v^j, \\
        A_{R^{-2},0.5}^j &= \frac{1}{2} (X_2^3 q_1 - X_1^3 q_2) \bigg[ 6 (\mathbf{N} \cdot \mathbf{v}) (\mathbf{N} \cdot \mathbf{r}) v^j - 3 \frac{G_{12} M}{r^3} (\mathbf{N} \cdot \mathbf{r})^2 r^j -2 r \dot{r} v^j + \frac{G_{12} M}{r} r^j \bigg],
\end{align}
\end{widetext}

\begin{widetext}
    \begin{align}
        A_{R^{-2},1}^j
&=
\frac{GM\dot r}{2}
\left[
q_2X_1(1-a\alpha_1)
+(3-a\alpha_2)q_1X_2
\right]N^j
\notag \\
&\quad
+\Bigg\{
-2\delta
\left(
q_2X_1-q_1X_2
\right)
(\mathbf N\cdot\mathbf r)
\notag \\
&\qquad
+\frac{GM}{2r}
\left[
q_2X_1
\left(
5+a\alpha_1-4(3+a\alpha_1)X_1
\right)
+
q_1X_2
\left(
5+a\alpha_2-4(3+a\alpha_2)X_2
\right)
\right]
(\mathbf N\cdot\mathbf r)
\notag \\
&\qquad
+
\left(
q_2X_1^4+q_1X_2^4
\right)
\bigg[
6(\mathbf N\cdot\mathbf r)(\mathbf N\cdot\mathbf v)^2
-4\frac{G_{12}M}{r^3}
(\mathbf N\cdot\mathbf r)^3
+\left(
\frac{2G_{12}M}{r}-v^2
\right)
(\mathbf N\cdot\mathbf r)
-2r\dot r(\mathbf N\cdot\mathbf v)
\bigg]
\Bigg\}v^j
\notag \\
&\quad
+\Bigg\{
-\dot\delta
\left(
q_2X_1-q_1X_2
\right)
(\mathbf N\cdot\mathbf r)
\notag \\
&\qquad
+\frac{GM}{2r}
\left[
(-3+a\alpha_1)q_2X_1
+
(-3+a\alpha_2)q_1X_2
\right]
(\mathbf N\cdot\mathbf v)
\notag \\
&\qquad
+\frac{GM\dot r}{2r^2}
\left[
(3+a\alpha_1)q_2X_1(2X_1-1)
+
(3+a\alpha_2)q_1X_2(2X_2-1)
\right]
(\mathbf N\cdot\mathbf r)
\notag \\
&\qquad
+
\left(
q_2X_1^4+q_1X_2^4
\right)
\bigg[
\frac{G_{12}M}{r}(\mathbf N\cdot\mathbf v)
+\frac{G_{12}M\dot r}{2r^2}(\mathbf N\cdot\mathbf r)
-6\frac{G_{12}M}{r^3}
(\mathbf N\cdot\mathbf r)^2(\mathbf N\cdot\mathbf v)
+3\frac{G_{12}M\dot r}{r^4}
(\mathbf N\cdot\mathbf r)^3
\bigg]
\Bigg\}r^j .
    \end{align}
\end{widetext}

\subsection{Gravitational waveform} \label{sec:grav_waveform}

The gravitational waveform is characterized by the transverse-traceless (TT) part of the metric perturbation $h_{jk}^{\text{TT}}$. We define the projection operator $P^{jk} = \delta^{jk} - N^j N^k$, and the spatial components $h^{jk}$ in the TT gauge are given by
\begin{align}\label{GWwaveformTT}
    h^{jk}_{\text{TT}} = (P^{jl} P^{km} - \frac{1}{2} P^{jk} P^{lm})h^{lm}.
\end{align}
For convenience, we apply the TT projector to the first two spatial indices of the EW moments. The remaining indices $(L=l_1\cdots l_\ell)$ label the multipolar order and are not affected by the TT projection. In the following expressions, the projection on the $(j,k)$ indices is understood whenever the subscript $({\rm TT})$ is used.

The PN accuracy required for each source moment follows directly from the wave-zone expansion (\ref{GWwaveform}). In the leading $R^{-1}$ field, $I^{jk}$, $I^{jk}_{l}$, $I^{jk}_{lm}$, and $I^{jk}_{lmn}$ first enter at relative \(0\), \(0.5\), \(1\), and \(1.5\)PN orders, respectively. Therefore, to construct the gravitational waveform through 1.5PN order, $I^{jk}$ and $I^{jk}_{l}$ must be known including their relative 1PN corrections, whereas the leading-order expressions of $I^{jk}_{lm}$ and $I^{jk}_{lmn}$ are sufficient. The required CM-frame moments are
\begin{widetext}
    \begin{align}
        I^{jk}_\text{TT} &= \eta \left[ 1 + \frac{1-3\nu}{2c^2} v^2 - \frac{G_{12} M (1-2\nu)}{2 c^2 r} \right] v^j v^k + \frac{G_{12} M \eta}{2 c^2 r} \dot{r} (\nu + \Gamma) (v^j n^k + n^j v^k) \notag \\
        &\quad - \frac{G_{12} M \eta}{r} \left\{ 1 + \frac{1}{c^2} \left[ \left( \frac{1}{4} + \frac{1}{2}\Gamma \right) v^2 - \left( \frac{3}{4} - \frac{3}{2}\nu \right) \dot{r}^2 - \left( \frac{3}{4} - \nu + \Gamma + \frac{G^2}{G_{12}^2} \Pi \right) \frac{G_{12} M}{r} \right] \right\} n^j n^k + \mathcal{O}(c^{-4}), \\
        I^{jk}_{l,\text{TT}} &= I^{jk}_{l,V,m} + I^{jk}_{l,V,\varphi} + I^{jk}_{l,V,A} + I^{jk}_{l,V,\text{LL}} + \mathcal{O}(c^{-4}), \label{IjklTT}\\
        I^{jk}_{lm,\text{TT}}
&=
\eta(1-3\nu)
\left(
v^jv^k-\frac{G_{12}M}{3r^3}r^jr^k
\right)r^lr^m
-\frac{G_{12} M\eta}{6r}\delta^{lm}r^jr^k
\notag\\
&\quad
+\frac{G_{12} M\eta}{6r}
\Bigg[
\frac{1}{2}
\left(
\delta^{jl}r^kr^m
+\delta^{jm}r^kr^l
+\delta^{kl}r^jr^m
+\delta^{km}r^jr^l
\right)
-r^2
\left(
\delta^{kl}\delta^{jm}
+\delta^{km}\delta^{jl}
\right)
\Bigg]
+\mathcal{O}(c^{-2}), \\
        I^{jk}_{lmn,\text{TT}}
&=
-\eta(X_1-X_2)(1-2\nu)
\left(
v^jv^k-\frac{G_{12}M}{4r^3}r^jr^k
\right)r^lr^mr^n
\notag\\
&\quad
+\frac{G_{12}M\eta(X_1-X_2)}{12r}
\left(
\delta^{mn}r^jr^kr^l
+\delta^{ln}r^jr^kr^m
+\delta^{lm}r^jr^kr^n
\right)
\notag\\
&\quad
-\frac{G_{12}M\eta(X_1-X_2)}{12r}
\Bigg[
\Big(
\delta^{jl}r^kr^mr^n
+\delta^{jm}r^kr^lr^n
+\delta^{jn}r^kr^lr^m
+\delta^{kl}r^jr^mr^n
+\delta^{km}r^jr^lr^n
+\delta^{kn}r^jr^lr^m
\Big)
\notag\\
&\qquad
-r^2
\Big(
\delta^{jl}\delta^{kn}r^m
+\delta^{jm}\delta^{kn}r^l
+\delta^{jl}\delta^{km}r^n
+\delta^{jn}\delta^{km}r^l
+\delta^{jm}\delta^{kl}r^n
+\delta^{jn}\delta^{kl}r^m
\Big)
\Bigg]
+\mathcal{O}(c^{-2}),
    \end{align}
\end{widetext}
The 1PN correction to the moment \(I^{jk}_{l}\) receives contributions from the compact matter source, the scalar-field stress-energy tensor, the electromagnetic-field stress-energy tensor, and the nonlinear gravitational source represented by the Landau-Lifshitz pseudotensor. We accordingly decompose it as in Eq. (\ref{IjklTT}), with the individual contributions given below
\begin{widetext}
    \begin{align}
        I^{jk}_{l,V,m} &= -\eta (X_1 - X_2) \Bigg\{ v^j v^k r^l \left[ 1 + \frac{1-5\nu}{2c^2} v^2 + \frac{G M}{c^2 r} (1-\nu)(3-\alpha_1\alpha_2) + \frac{3\nu G_{12} M}{2 c^2 r} \right] \notag \\
        &\quad + \frac{G_{12} M \dot{r}}{2 c^2 r^2} \Big[ (v^j r^k r^l + v^k r^j r^l) \left( \nu + \Gamma - \frac{G}{G_{12}}(1-\nu)(3-\alpha_1\alpha_2) \right) - v^l r^j r^k \left( \Gamma - \frac{G}{G_{12}}(1-\nu)(3- \alpha_1\alpha_2) \right) \Big] \notag \\
        &\quad + \frac{G_{12} M}{2 r^3} r^j r^k r^l \Bigg[ -1 + \frac{1}{c^2} \Bigg( \frac{\nu-\Gamma}{2} v^2 + \frac{G_{12} M}{r} \left( \Gamma-\nu + \frac{G^2}{G_{12}^2}\Pi - \frac{G}{G_{12}}(1-\nu)(3-\alpha_1\alpha_2) \right) \Bigg) \Bigg] \Bigg\}, \\
        I^{jk}_{l,V,\varphi} &= -\eta (X_1 - X_2) \frac{G M \alpha_1 \alpha_2}{12 c^2} \Bigg\{ \frac{2 G_{12} M + r \dot{r}^2 - r v^2}{2 r^2} (\delta_{jl} r^k + \delta_{kl} r^j) \notag \\
        &\quad - \dot{r} (\delta_{jl} v^k + \delta_{kl} v^j) - \left[ \frac{2 G_{12} M}{r^4} - \frac{3(1-3\nu)\dot{r}^2}{r^3} + \frac{(1-3\nu)v^2}{r^3} \right] r^j r^k r^l \notag \\
        &\quad - \frac{2(1-3\nu)\dot{r}}{r^2} (v^j r^k r^l + v^k r^j r^l) - \frac{2\dot{r}}{r^2} v^l r^j r^k + \frac{2(1-3\nu)}{r} v^j v^k r^l + \frac{2}{r} (v^j v^l r^k + v^k v^l r^j) \Bigg\}, \\
        I^{jk}_{l,V,A} &= \frac{q_1 q_2 (X_1 - X_2)}{2c^2} \Bigg\{ \frac{4 G_{12} M + r(v^2 - \dot{r}^2)}{12 r^2} (\delta_{kl} r^j + \delta_{jl} r^k) \notag \\
        &\quad + \frac{2 G_{12} M (-2 + 3\nu) + (1 - 3\nu) r (3\dot{r}^2 - v^2)}{6 r^4} r^j r^k r^l + \frac{2 - 3\nu}{3 r} v^j v^k r^l \notag \\
        &\quad - \frac{\dot{r}}{3} (\delta_{kl} v^j + \delta_{jl} v^k) - \frac{\dot{r}}{6 r^2} (r^k v^j + r^j v^k) r^l - \frac{(1 - 3\nu)\dot{r}}{3 r^2} r^j r^k v^l + \frac{1}{6 r} (r^k v^j + r^j v^k) v^l \Bigg\}, \\
        I^{jk}_{l,V,\text{LL}} &= \frac{\eta M (X_1 - X_2) G}{2c^2} \Bigg\{ \frac{22 G_{12} M + r(v^2 - \dot{r}^2)}{12 r^2} (\delta_{kl} r^j + \delta_{jl} r^k) \notag \\
        &\quad - \frac{2 G_{12} M (11 - 12\nu) + (1 - 3\nu) r (3\dot{r}^2 - v^2)}{6 r^4} r^j r^k r^l + \frac{11 - 21\nu}{3 r} v^j v^k r^l \notag \\
        &\quad - \frac{11}{6} \dot{r} (\delta_{kl} v^j + \delta_{jl} v^k) + \frac{(-5 + 9\nu)\dot{r}}{3 r^2} (r^k v^j + r^j v^k) r^l + \frac{(1 - 12\nu)\dot{r}}{3 r^2} r^j r^k v^l + \frac{5}{3 r} (r^k v^j + r^j v^k) v^l \Bigg\}.
\end{align}
\end{widetext}
where $\eta = m_1 m_2 / M$ is the reduced mass, $\nu = \eta/M$ is the symmetric mass ratio, $X_A = m_A / M$ are the mass fractions, and $G_{12} = G(1 + \alpha_1 \alpha_2 - q_1 q_2/G M \eta)$ represents the effective 1PN gravitational interaction. The time derivative $d/dt$ acts on the entire bracketed expression. Note that all tensors on the right-hand sides of the above equations are understood to be projected on their $j,k$ indices according to Eq. (\ref{GWwaveformTT}).

Substituting these multipole moments into the standard expansion for the gravitational field in the wave zone, the metric perturbation $h_{jk}^{\text{TT}}$ can be decomposed into terms of different orders in $1/R$ and $1/c$:
\begin{align}
    h_{jk}^{\text{TT}} &= \frac{4 G}{c^4 R} [h^{jk}_{R^{-1}, 0} + c^{-1} h^{jk}_{R^{-1}, 0.5} + c^{-2} h^{jk}_{R^{-1}, 1} \notag \\
    &\quad + c^{-3} h^{jk}_{R^{-1}, 1.5}]_{\text{TT}} \notag \\
    &\quad + \frac{4 G}{c^4 R^2} [h^{jk}_{R^{-2}, 0} + c^{-1} h^{jk}_{R^{-2}, 0.5} + c^{-2} h^{jk}_{R^{-2}, 1}]_{\text{TT}} \notag \\
    &\quad + \mathcal{O}(c^{-7}).
\end{align}
For the computation of the instantaneous gravitational angular momentum flux through relative 1PN order, only the radiative \(\mc{O}(R^{-1})\) part of the TT metric perturbation is required through 1PN accuracy. Nevertheless, we also present the \(\mc{O}(R^{-1})\) waveform through 1.5PN order for completeness:
\begin{widetext}
    \begin{align}
        h^{jk}_{R^{-1},0,\text{TT}} &= \eta \bigg( v^j v^k - \frac{G_{12} M}{r} n^j n^k \bigg)_{\text{TT}}, \\
        h^{jk}_{R^{-1},0.5,\text{TT}} &= - \frac{\eta (X_1 - X_2)}{2} \bigg[ 2 (\mathbf{N} \cdot \mathbf{v}) v^j v^k - 3 \frac{G_{12} M}{r^3} (\mathbf{N} \cdot \mathbf{r}) (v^j r^k + r^j v^k) \notag \\
        &\quad - \frac{G_{12} M}{r^3} (\mathbf{N} \cdot \mathbf{v}) r^j r^k + 3 \frac{G_{12} M \dot{r}}{r^4} (\mathbf{N} \cdot \mathbf{r}) r^j r^k \bigg]_{\text{TT}}, \\
        h^{jk}_{R^{-1},1,\text{TT}}
&=
\Bigg\{
\eta
\Bigg[
\frac{1-3\nu}{2}v^2
+\left(\nu-\frac{2}{3}\right)\frac{G_{12}M}{r}
+(1-3\nu)(\mathbf N\cdot\mathbf v)^2
-\frac{7}{3}(1-3\nu)\frac{G_{12}M}{r^3}
(\mathbf N\cdot\mathbf r)^2
\Bigg]v^jv^k
\notag\\
&\quad
+\frac{G_{12}M\eta}{r^2}
\Bigg[
\frac{\dot r}{6}
\left(
1+3\nu+3\Gamma
\right)
+(1-3\nu)
\left(
\frac{5\dot r}{2r^2}(\mathbf N\cdot\mathbf r)^2
-\frac{8}{3r}(\mathbf N\cdot\mathbf r)(\mathbf N\cdot\mathbf v)
\right)
\Bigg](v^jr^k+r^jv^k)
\notag\\
&\quad
+\frac{G_{12} M \eta}{r^3}
\Bigg[
\frac{1}{6}
\bigg[
\frac{G_{12}M}{r}\left(5-6\nu+6\Gamma+6\frac{G^2}{G_{12}^2}\Pi\right)
-
\big(
(-3+9\nu)\dot r^2
+(1+3\Gamma)v^2
\big)
\bigg]
\notag\\
&\qquad
+(1-3\nu)
\Bigg[
\frac{2\dot r}{r}(\mathbf N\cdot\mathbf r)(\mathbf N\cdot\mathbf v)
-\frac{1}{3}(\mathbf N\cdot\mathbf v)^2
+\frac{1}{6r^2}
\left(
7\frac{G_{12}M}{r}
+3(-5\dot r^2+v^2)
\right)
(\mathbf N\cdot\mathbf r)^2
\Bigg]
\Bigg]r^jr^k
\Bigg\}_{\text{TT}}, \\
        h^{jk}_{R^{-1},1.5,\text{TT}}
&=
\Big\{
\mathcal H_{rr}\,r^jr^k
+\mathcal H_{rv}\left(r^jv^k+v^jr^k\right)
+\mathcal H_{vv}\,v^jv^k
\Big\}_{\text{TT}},
\end{align}
\end{widetext}
with the coefficients $\mathcal H_{rr}$, $\mathcal H_{rv}$ and $\mathcal H_{vv}$ deferred to Appendix~\ref{app:grav_moment_coeff} for compactness. And the $1/R^2$ terms represent the next-to-leading order corrections of gravitational waveforms:
\begin{widetext}
    \begin{align}
        h^{jk}_{R^{-2},0,\text{TT}} &= - \eta (X_1 - X_2) (\mathbf{N} \cdot \mathbf{r}) \left( v^j v^k - \frac{G_{12} M}{2 r^3} r^j r^k \right)_{\text{TT}}, \\
        h^{jk}_{R^{-2},0.5,\text{TT}}
&=
\Bigg\{
\eta(1-3\nu)
\left[
3(\mathbf N\cdot\mathbf v)(\mathbf N\cdot\mathbf r)
-r\dot r
\right]v^jv^k
+\frac{\eta G_{12}M}{2r}
\left[
1-4\nu
-4(1-3\nu)
\frac{(\mathbf N\cdot\mathbf r)^2}{r^2}
\right]
\left(
v^jr^k+r^jv^k
\right)
\notag\\
&\quad
+\frac{\eta G_{12}M}{2r^4}
\left[
\nu r^2\dot r
-(1-3\nu)
\left(
2r(\mathbf N\cdot\mathbf r)(\mathbf N\cdot\mathbf v)
-3\dot r(\mathbf N\cdot\mathbf r)^2
\right)
\right]
r^jr^k
\Bigg\}_{\text{TT}}, \\
        h^{jk}_{R^{-2},1,\text{TT}}
&=
\Bigg\{
\mathcal H^{(2)}_{rr}\,r^jr^k
+\mathcal H^{(2)}_{rv}\left(r^jv^k+v^jr^k\right)
+\mathcal H^{(2)}_{vv}\,v^jv^k
\Bigg\}_{\text{TT}},
\end{align}
\end{widetext}
with the coefficients $\mathcal H_{rr}^{(2)}$, $\mathcal H_{rv}^{(2)}$ and $\mathcal H_{vv}^{(2)}$ deferred to Appendix~\ref{app:grav_moment_coeff} for compactness.

\section{The Angular Momentum Flux} \label{sec:flux_evo}

The evolution of circular orbit consists of only one degree of freedom, the effective one-body radius $r$. After selecting an appropriate CM system, the energy loss can be computed directly using energy flux defined by \cite{Khalil:2018aaj, Damour:1992we, Astefanesei:2019pfq, Poisson_Will_2014, Blanchet:2002av}

\begin{align}
    \mathcal{F}_{\varphi} &= \frac{c^3 R^2}{4 \pi G} \int d\Omega \dot{\varphi}^2, \\
    \mathcal{F}_{A} &= \frac{R^2}{4 \pi c} \int d\Omega \left( \dot{A}^k \dot{A}_k - N_j N_k \dot{A}^j \dot{A}^k \right), \\
    \mathcal{F}_{\text{LL}} &= \frac{c^3 R^2}{32 \pi G} \int d\Omega \left( \dot{h}^{jk}_{\text{TT}} \dot{h}^{\text{TT}}_{jk} \right).
\end{align}
For quasicircular inspirals, the secular orbital evolution follows from the energy-balance equation
\begin{equation}
    \langle \frac{d E}{dt} \rangle = - \langle \mathcal{F} \rangle,
\end{equation}
where $\langle ... \rangle$ denotes an orbital average over a period. Once the energy flux $\mathcal{F}$ on the right-hand side has been evaluated, the equation above can determine the evolution of the variable $r$.

However, to extend the quasicircular orbit to the noncircular or quasielliptical case, an additional degree of freedom should be taken into account to determine the orbital evolution.
Here, we suppose the binary BH system is always in the $x$-$y$ plane, then $z$-component of angular momentum and flux can determine the orbital evolution. For nonspinning binaries, the relative acceleration in Eq.~\eqref{eq:EoM} is a linear combination of \(\mathbf n\) and \(\mathbf v\), with scalar coefficients depending only on \(r\), \(v^2\), and \(\dot r\). Therefore, if the initial position and velocity lie in a given plane, the acceleration remains in the same plane and no out-of-plane motion is generated. Equivalently, the direction of \(\mathbf{r}\times\mathbf{v}\), which is normal to the orbital plane, remains fixed. We may therefore choose the orbital plane to be the $x$-$y$ plane without loss of generality.  Then, one can introduce the angular momentum flux $\mathcal{J}^{jk}$ to describe the loss of total angular momentum $L^z$ of binary BH system, also by applying an additional balance equation

\begin{equation}
    \langle \frac{d L^z}{dt} \rangle = - \langle \mathcal{J}^{z} \rangle,
\end{equation}
where the $z$-component $L^z$ of total angular momentum of the binary BH system is defined by

\begin{equation}
    L^z = \frac{1}{c} \int_\mathcal{M} \left( x \mu^{y0} - y \mu^{x0} \right) d^3x, \label{eq:Def_Lz}
\end{equation}
the angular momentum flux in the vectorial form is given by

\begin{equation}
    \mathcal{J}^j = \frac{1}{2} \epsilon^j_{kl} \mathcal{J}^{kl},
\end{equation}
and the tensorial angular momentum flux consists of scalar, electromagnetic and gravitational contributions,

\begin{equation}
    \mathcal{J}^{kl} = \mathcal{J}^{kl}_\varphi + \mathcal{J}^{kl}_A + \mathcal{J}^{kl}_{\text{TT}},
\end{equation}
their explicit forms read
\begin{widetext}
    \begin{align}
    \mathcal{J}^{jk}_\varphi &= -\frac{c^3}{4\pi G} \int R^3 \dot{\varphi} \left( N^j \eth^k_T \varphi - N^k \eth^j_T \varphi \right) d\Omega, \label{eq:Def_scalar_flux} \\
    \mathcal{J}^{jk}_A &= \frac{1}{4 \pi c} \int \Bigg\{ R^3 \Bigg[ N^j \Big( \dot{A}^0 \eth^k_T A^0 - \dot{A}^l \eth^k_T A_l + \dot{A}^l \eth^T_l A^k - \dot{A}^k \eth^T_l A^l - h^{00} \dot{A}^k N_l \dot{A}^l \Big) \notag \\
    &\quad\ - N^k \Big( \dot{A}^0 \eth^j_T A^0 - \dot{A}^l \eth^j_T A_l + \dot{A}^l \eth^T_l A^j - \dot{A}^j \eth^T_l A^l - h^{00} \dot{A}^j N_l \dot{A}^l \Big) \Bigg] \notag \\
    & + R^2 \Bigg[ 2(N_l A^l) (N^j \dot{A}^k - N^k \dot{A}^j) - (N_l \dot{A}^l) (N^j A^k - N^k A^j) \Bigg] \Bigg\} d\Omega, \label{eq:Def_EM_flux} \\
    \mathcal{J}^{jk}_{\text{LL}} &= \frac{c^3}{16\pi G} \int R^2 \Bigg[ h_{\text{TT}}^{jp} \dot{h}_{\text{TT}}^{kp} - h_{\text{TT}}^{kp} \dot{h}_{\text{TT}}^{jp} - \frac{1}{2} \dot{h}_{\text{TT}}^{pq} (x^j \partial^k - x^k \partial^j) h^{\text{TT}}_{pq} \Bigg] d\Omega, \label{eq:Def_grav_flux}
    \end{align}
\end{widetext}
where $\eth^j_T = P^j_k \eth^k$ denotes the transverse derivative. And the spatial partial derivative operator $\eth^j$ acting on functions of form $f(\tau,R,\mathbf{N})$ represents that it takes derivative with respect to the explicit spatial dependence $R$ or $\mathbf{N}$ and neglects the implicit spatial dependence in the retarded time $\tau = t-R/c$. The detailed explanation and calculation of the angular momentum fluxes presented above are deferred to Appendix \ref{app:ang_moment_flux}.

Evaluating Eq.~\eqref{eq:Def_Lz} through 1PN order yields the total angular momentum
\begin{align}
    L^z &= \eta h \Bigg[ 1 + \frac{v^2}{2c^2}(1 - 3\nu) + \frac{G_{12} M}{c^2 r} \Gamma \Bigg] + \mathcal{O}(c^{-4}),
\end{align}
where $h = |\mathbf{r} \times \mathbf{v}|$ denotes the Newtonian specific orbital angular momentum of the relative motion.

Now that we have established the explicit formulas for computing angular momentum fluxes and computed the total angular momentum of binary BH system, the next step is to obtain the explicit result of angular momentum flux up to 1PN order.

\subsection{Scalar angular momentum flux}

The scalar angular momentum flux $\mathcal{J}^{jk}_\varphi$ can be computed using Eq.~\eqref{eq:Def_scalar_flux}. The quadrupole contribution is given by
\begin{multline}
\mathcal J^{jk}_{\varphi,\mathrm{quad}}
 =\frac{4G}{15c^5}\frac{G_{12}M\eta^2}{r^2}
 \left(2\frac{G_{12}M}{r}+2v^2-3\dot r^2\right) \\
{}\times(\alpha_2X_1+\alpha_1X_2)^2
 (n^jv^k-v^jn^k).
\end{multline}
and the dipole contribution is
\begin{multline}\label{eq:scalar_leading_0.5}
\mathcal J^{jk}_{\varphi,\mathrm{dipole}}
 =\frac{G}{3c^3}\frac{G_{12}M\eta^2}{r^2}
 \Bigg[(\alpha_1-\alpha_2)^2
 +j^{\varphi}_{\dot r^2}\frac{\dot r^2}{c^2} \\ 
{}+j^{\varphi}_{v^2}\frac{v^2}{c^2}
 +j^{\varphi}_{1/r}\frac{G_{12}M}{c^2r}\Bigg]
 (n^jv^k-v^jn^k).
\end{multline}
with the coefficients
    \begin{align}
j^{\varphi}_{\dot r^2}
&=
-3(\alpha_1-\alpha_2)
\Bigg\{
\notag\\
&\quad
\frac{GX_1}{G_{12}}
\big(\alpha_2+\alpha_1\alpha_2^2+\alpha_1\beta_2\big)
\notag\\
&\qquad-\frac{GX_2}{G_{12}}
\big(\alpha_1+\alpha_1^2\alpha_2+\alpha_2\beta_1\big)
\notag\\
&\qquad
-\frac{a q_1q_2}{G_{12}M\eta}(X_1-X_2)
+\frac{1}{2}(\alpha_1+\alpha_2)(X_1-X_2)
\Bigg\},
\\[1ex]
j^{\varphi}_{v^2}
&=
(\alpha_1-\alpha_2)
\Bigg\{
\notag\\
&\quad
\frac{GX_1}{G_{12}}
\big(\alpha_2+\alpha_1\alpha_2^2+\alpha_1\beta_2\big)
\notag\\
&\qquad-\frac{GX_2}{G_{12}}
\big(\alpha_1+\alpha_1^2\alpha_2+\alpha_2\beta_1\big)
\notag\\
&\qquad
-\frac{a q_1q_2}{G_{12}M\eta}(X_1-X_2)
\notag\\
&\qquad
+\frac{\alpha_1-\alpha_2}{20}
\Big[
1+(X_1-X_2)^2+10\Gamma
\Big]
\notag\\
&\qquad
+\frac{9}{10}(\alpha_1+\alpha_2)(X_1-X_2)
\Bigg\},
\\[1ex]
j^{\varphi}_{1/r}
&=
(\alpha_1-\alpha_2)
\Bigg\{
\notag\\
&\quad
\frac{GX_1}{G_{12}}
\big(\alpha_2+\alpha_1\alpha_2^2+\alpha_1\beta_2\big)
\notag\\
&\qquad-\frac{GX_2}{G_{12}}
\big(\alpha_1+\alpha_1^2\alpha_2+\alpha_2\beta_1\big)
\notag\\
&\qquad
-\frac{a q_1q_2}{G_{12}M\eta}(X_1-X_2)
\notag\\
&\qquad
-(\alpha_1-\alpha_2)
\Bigg[
\Gamma
+\frac{G^2}{G_{12}^2}\Pi
+\frac{19+29(X_1-X_2)^2}{40}
\Bigg]
\notag\\
&\qquad
-\frac{3}{10}(\alpha_1+\alpha_2)(X_1-X_2)
\Bigg\}.
\end{align}

\subsection{Electromagnetic angular momentum flux}

The electromagnetic angular momentum flux $\mathcal{J}^{jk}_{A}$ can be computed using Eq.~\eqref{eq:Def_EM_flux}. The quadrupole contribution is given by
\begin{multline}
\mathcal J^{jk}_{A,\mathrm{quad}}
 =\frac{2}{5c^5}\frac{G_{12}M\eta^2}{r^2}
 \left(2\frac{G_{12}M}{r}-3\dot r^2+2v^2\right) \\
{}\times\left(\frac{q_2}{m_2}X_1+\frac{q_1}{m_1}X_2\right)^2
 (n^jv^k-v^jn^k),
\end{multline}
and the dipole contribution is
\begin{multline}\label{eq:EM_leading_0.5}
\mathcal J^{jk}_{A,\mathrm{dipole}}
 =\frac{2}{3c^3}\frac{G_{12}M\eta^2}{r^2}
 \Bigg[\left(\frac{q_2}{m_2}-\frac{q_1}{m_1}\right)^2
 +j^A_{\dot r^2}\frac{\dot r^2}{c^2} \\
{}+j^A_{v^2}\frac{v^2}{c^2}
 +j^A_{1/r}\frac{G_{12}M}{c^2r}\Bigg]
 (n^jv^k-v^jn^k).
\end{multline}
with the coefficients
\begin{align}
j^A_{\dot r^2}
&=
\left(
\frac{q_2}{m_2}
-\frac{q_1}{m_1}
\right)
\Bigg[
\notag\\
&\quad
3\frac{G}{G_{12}}
\left[
a\left(
\alpha_1\frac{q_2}{m_2}
-\alpha_2\frac{q_1}{m_1}
\right)
+
\left(
\frac{q_2}{m_2}
-\frac{q_1}{m_1}
\right)
\right]
\notag\\
&\quad
+\frac{3\nu}{2}
\left(
\frac{q_1}{m_2}
-\frac{q_2}{m_1}
\right)
\notag\\
&\quad
+\frac{3\eta^2}{10}
\left(
\frac{q_2}{m_2^3}
-\frac{q_1}{m_1^3}
\right)
\Bigg],
\\[1ex]
j^A_{v^2}
&=
\left(
\frac{q_2}{m_2}
-\frac{q_1}{m_1}
\right)
\Bigg[
\notag\\
&\quad
-\frac{G}{G_{12}}
\left[
a\left(
\alpha_1\frac{q_2}{m_2}
-\alpha_2\frac{q_1}{m_1}
\right)
+
\left(
\frac{q_2}{m_2}
-\frac{q_1}{m_1}
\right)
\right]
\notag\\
&\quad
-\frac{\nu}{2}
\left(
3\frac{q_1}{m_2}
-2\frac{q_2}{m_2}
+2\frac{q_1}{m_1}
-3\frac{q_2}{m_1}
\right)
\notag\\
&\qquad
+\frac{1}{10}\frac{q_1}{m_1}
\left(
5-5\Gamma
-3\frac{\eta^2}{m_1^2}
\right)
\notag\\
&\qquad
+\frac{1}{10}\frac{q_2}{m_2}
\left(
-5+5\Gamma
+3\frac{\eta^2}{m_2^2}
\right)
\Bigg],
\\[1ex]
j^A_{1/r}
&=
-\left(
\frac{q_2}{m_2}
-\frac{q_1}{m_1}
\right)
\Bigg[
\notag\\
&\quad
\frac{G}{G_{12}}
\left[
a\left(
\alpha_1\frac{q_2}{m_2}
-\alpha_2\frac{q_1}{m_1}
\right)
+
\left(
\frac{q_2}{m_2}
-\frac{q_1}{m_1}
\right)
\right]
\notag\\
&\quad
+\frac{G^2}{G_{12}^2}
\left(
\frac{q_2}{m_2}
-\frac{q_1}{m_1}
\right)\Pi
\notag\\
&\qquad
+\frac{1}{10}
\Bigg(
-5\nu\frac{3q_1+q_2}{m_2}
+5\nu\frac{q_1+3q_2}{m_1}
\notag\\
&\qquad
+10\Gamma
\left(
\frac{q_2}{m_2}
-\frac{q_1}{m_1}
\right)
\notag\\
&\qquad
+7\eta^2
\left(
\frac{q_2}{m_2^3}
-\frac{q_1}{m_1^3}
\right)
\Bigg)
\Bigg].
\end{align}

\subsection{Gravitational angular momentum flux}

The tensor gravitational-wave angular momentum flux $\mathcal{J}^{jk}_{\text{LL}}$ can be computed using Eq.~\eqref{eq:Def_grav_flux}.  The quadrupole contribution is
\begin{multline}
\mathcal J^{jk}_{\mathrm{LL},\mathrm{quad}}
 =\frac{8G}{5c^5}\frac{G_{12}M\eta^2}{r^2}
 \Bigg[2\frac{G_{12}M}{r}-3\dot r^2+2v^2 \\
{}+\frac{1}{c^2}j^{\mathrm{LL}}_{\mathrm{quad}}\Bigg]
 (n^jv^k-v^jn^k),
\end{multline}
where
\begin{align}
j^{\text{LL}}_{\text{quad}}
&=\left(-\frac{82}{21}+\frac{87}{14}\nu-\frac72\Gamma
-4\frac{G^2}{G_{12}^2}\Pi\right)
\left(\frac{G_{12}M}{r}\right)^2 \notag\\
&\quad+\left(-\frac{32}{7}+\frac{75}{7}\nu
-2\frac{G^2}{G_{12}^2}\Pi\right)
\frac{G_{12}M}{r}v^2 \notag\\
&\quad+\left(\frac{118}{21}-\frac{159}{14}\nu+\Gamma
+4\frac{G^2}{G_{12}^2}\Pi\right)
\frac{G_{12}M}{r}\dot r^2 \notag\\
&\quad+\frac{46-117\nu+21\Gamma}{42}v^4
+\frac{19}{14}(-1+3\nu)\dot r^2v^2 \notag\\
&\quad+\frac57(1-3\nu)\dot r^4.
\end{align}
The octupole contribution is
\begin{multline}
\mathcal J^{jk}_{\mathrm{LL},\mathrm{oct}}
 =\frac{8G}{5c^7}\frac{G_{12}M\eta^2}{r^2}
 \frac{(X_1-X_2)^2}{84} \\
{}\times\Bigg[56\left(\frac{G_{12}M}{r}\right)^2
 +4\frac{G_{12}M}{r}(-79\dot r^2+80v^2) \\
{}+225\dot r^4-330\dot r^2v^2+89v^4\Bigg]
 (n^jv^k-v^jn^k).
\end{multline}

\subsection{Discussion}

In this subsection, we discuss the main features of the angular-momentum
flux in EMd theory and assess the consistency of our results in three
relevant limits: the quasicircular, GR, and dipole-suppression limits.

A common structural property of all radiative channels is that the instantaneous flux is proportional to ($n^jv^k-v^jn^k$). Equivalently, the vectorial flux is parallel to \(\bm n\times\bm v\), and hence to the orbital angular momentum $\mathbf{L}$. Therefore, for a nonspinning planar binary, radiation changes the magnitude of the orbital angular momentum but does not tilt the orbital plane at the order considered here.

\begin{table*}[t]
    \caption{
        Leading multipolar structure of the angular momentum flux.
    }
    \label{tab:flux_hierarchy}
    \begin{ruledtabular}
    \begin{tabular}{cccc}
        Channel & Leading multipole & Scaling & Suppression condition \\
        \hline
        Scalar & Dipole & \(c^{-3}\) &
        \(\alpha_1=\alpha_2\) \\
        Electromagnetic & Dipole & \(c^{-3}\) &
        \(q_1/m_1=q_2/m_2\) \\
        Scalar & Quadrupole & \(c^{-5}\) & --- \\
        Electromagnetic & Quadrupole & \(c^{-5}\) & --- \\
        Tensor & Quadrupole & \(c^{-5}\) & --- \\
        Tensor & Octupole & \(c^{-7}\) & --- \\
    \end{tabular}
    \end{ruledtabular}
\end{table*}

As summarized in Table~\ref{tab:flux_hierarchy}, the scalar and electromagnetic channels begin with dipole radiation at \(\mathcal O(c^{-3})\), while their relative 1PN corrections and leading quadrupole contributions enter at \(\mathcal O(c^{-5})\). The gravitational channel begins with quadrupole radiation at \(\mathcal O(c^{-5})\), with its relative 1PN correction and leading octupole contribution entering at \(\mathcal O(c^{-7})\). Accordingly, the scalar and electromagnetic dipole channels appear formally at \(-1\)PN order relative to the leading gravitational quadrupole radiation.

As the first consistency check, in the quasicircular limit, i.e., $\dot{r} = 0$, the vectorial angular momentum flux $\mathcal{J}^j$ must satisfy the relation $\Omega \mathcal{J}^z = \mathcal{F}$, where $\Omega = v/r$ is the orbital angular frequency. Up to 1PN order, the orbital angular frequency $\Omega$ reads
\begin{align}
    \Omega^2
    &=
    \frac{G_{12}M}{r^3}
    \left[
        1
        +
        \left(
            -\frac12
            +\frac32\nu
            -\frac12\Gamma
            -\frac{G^2}{G_{12}^2}\Pi
        \right)
        \frac{G_{12}M}{c^2r}
    \right]
    \notag \\
    &\quad
    +\mathcal O(c^{-4}) . \label{eq:OAF}
\end{align}
Using Eq.~\eqref{eq:OAF}, we impose the quasicircular limit $\dot{r}=0$ and compare $\Omega \mathcal{J}^z$ with the energy flux $\mathcal{F}$ of Ref.~\cite{Khalil:2018aaj}. The two results agree through 1PN order.

As a second consistency check, we consider the GR limit by setting
the scalar couplings and electric charges to zero,
\begin{equation}
    \alpha_A=\beta_A=q_A=0.
\end{equation}
Under this limit, the effective two-body coupling reduces to the
Newtonian gravitational constant,
\begin{equation}
    G_{12}
    =
    G\left(
        1+\alpha_1\alpha_2
        -\frac{q_1q_2}{GM\eta}
    \right)
    \longrightarrow G.
\end{equation}
The two combinations entering the 1PN dynamics and the gravitational
flux consequently reduce to
\begin{align}
    \Gamma
    &=
    \frac{3-\alpha_1\alpha_2}
    {1+\alpha_1\alpha_2-q_1q_2/(GM\eta)}
    +\nu
    \longrightarrow 3+\nu,
    \\
    \Pi
    &\longrightarrow 1.
\end{align}
Moreover, all scalar and electromagnetic waveforms, and hence their corresponding angular momentum fluxes, vanish in this limit. The total flux therefore reduces entirely to the tensor gravitational contribution. After these substitutions, both the quadrupole and octupole gravitational contributions agree with the generic-orbit GR result~\cite{Junker:1992kle,Jain:2024lie}.

As the final consistency check, we consider the dipole-suppression limits summarized in Table~\ref{tab:flux_hierarchy}. The scalar dipole channel is suppressed when \(\alpha_1=\alpha_2\), since the leading dipolar waveform in Eq.~\eqref{eq:scalar_leading_0.5} is proportional to \(\alpha_1-\alpha_2\). Moreover, the coefficients of all relative 1PN corrections to the scalar dipole flux contain the same overall factor, so the complete scalar dipole contribution vanishes through the order considered here. Similarly, the electromagnetic dipole channel is suppressed when \(q_1/m_1=q_2/m_2\), because the leading dipolar waveform in Eq.~\eqref{eq:EM_leading_0.5}, as well as all of its relative 1PN corrections, is proportional to the difference in the charge-to-mass ratios. Our results therefore exhibit the expected suppression of both dipole channels.

\section{Conclusion} \label{sec:conclu}

In this work we have calculated the radiative contributions needed to describe noncircular nonspinning black hole binaries in EMd theory beyond the quasicircular model. Starting from the Einstein-frame formulation of the theory, we treated the binary as a system of skeletonized compact objects carrying scalar and electric charges, and used the DIRE approach to construct the scalar, electromagnetic, and gravitational radiation fields in the wave zone.

In addition to the radiative \(R^{-1}\) fields required for the angular momentum flux at null infinity, we obtained the next-to-leading \(R^{-2}\) wave-zone contributions. These terms describe finite-distance corrections and do not enter the
null-infinity flux. They constitute a separate extension of the wave-zone field calculation and may be useful in future studies of finite-radius waveform extraction.

The main result of this paper is the instantaneous angular momentum flux for generic noncircular orbits up to relative first post-Newtonian order. The flux receives contributions from all three radiative channels of the theory: scalar radiation, electromagnetic radiation, and tensor gravitational radiation. The noncircular case is qualitatively richer than the circular one because radial motion activates radiative structures that are absent or degenerate in the quasicircular limit.

Several independent limits provide consistency checks on our angular momentum flux. In the quasicircular limit, $\dot r=0$, the flux satisfies the balance relation $\Omega \mathcal{J}^{z}=\mathcal{F}$ and agrees with the energy-flux result of Ref.~\cite{Khalil:2018aaj} through relative 1PN order. In the GR limit, $\alpha_A=\beta_A=q_A=0$, the scalar and electromagnetic contributions vanish, while the remaining tensor contribution reduces to the known generic-orbit GR result \cite{Jain:2024lie,Junker:1992kle}. Finally, the expected dipole-suppression limits are recovered through the order considered here: the scalar dipole contribution vanishes for $\alpha_1=\alpha_2$, whereas the electromagnetic dipole contribution vanishes for $q_1/m_1=q_2/m_2$.

The results derived here also prepare the ground for orbit-averaged eccentric evolution, waveform phasing, and more systematic tests of charged black hole binaries in future gravitational-wave observations.

\begin{acknowledgments}
The work is in part supported by NSFC Grant No.12205104 and the student's research project of SCUT.
\end{acknowledgments}

\appendix

\section{Wave-zone contributions to the wave-zone field point} \label{app:wavezone_contribution}

In this appendix, we calculate the wave-zone contributions to the wave-zone solutions up to relative first post-Newtonian order and show that they can be neglected in the angular momentum flux calculation.

The solutions in the wave zone can be decomposed into two parts: the contribution from the near zone and the contribution from the wave zone. The near-zone contribution has been discussed in Sec.~\ref{sec:formal_solution}. Here we consider the contribution from the wave zone. Following the wave-zone integration method of Refs.~\cite{Will:1996zj, Shiralilou:2021mfl}, the leading wave-zone scalar source is
\begin{equation}
    \mu_\mathcal{W}
    =
    -\frac{aQ^2}{4\pi R'^4},
    \qquad
    Q=q_1+q_2 ,
\end{equation}
where \(R'=|\mathbf{x}'|\). Equivalently, in the form we require,
\begin{equation}
    \mu_\mathcal{W}
    =
    \frac{1}{4\pi}
    \frac{f_\ell(\tau) N'^{\langle L \rangle}}{R'^4},
    \qquad
    \ell
    =
    0,
    \qquad
    f_0
    =
    - a Q^2,
\end{equation}
where $\mathbf{N'} = \mathbf{x'}/R'$ denotes the unit direction vector pointing to source point in the wave zone. $N'^{\langle L \rangle}$ is an abbreviation of collection of unit direction vectors $N'^{\langle j_1 j_2,...,j_\ell \rangle}$, and the bracket $\langle \rangle$ in the superscript represents $N'^L$ in the STF form.

The corresponding wave-zone contribution to the scalar field is \cite{Will:1996zj, Shiralilou:2021mfl}
\begin{equation}
\begin{aligned}
    \varphi_\mathcal{W}
    &=
    \frac{G}{c^4}
    \frac{N^{\langle L \rangle}}{R}
    \Bigg[
    \int_0^{\mathcal R}
    f_\ell(\tau-2s/c) A(s,R)\,ds
    \\
    &\qquad
    +
    \int_{\mathcal R}^{\infty}
    f_\ell(\tau-2s/c) B(s,R)\,ds
    \Bigg] .
\end{aligned}
\end{equation}
where
\begin{align}
    A(s,R)
    =
    \int_{\mathcal R}^{R+s}
    P_\ell(\xi)\,p^{-(n-1)}\,dp, \\
    B(s,R)
    =
    \int_s^{R+s}
    P_\ell(\xi)\,p^{-(n-1)}\,dp,
\end{align}
with
\begin{align}
    \xi
    =
    \frac{R+2s}{R}
    -
    \frac{2s(R+s)}{Rp}.
\end{align}

For \(\ell=0\) and \(n=4\), discarding the boundary-dependent term, the finite wave-zone contribution is
\begin{align}
    \varphi_\mathcal{W}
    =
    \frac{aGQ^2}{2c^4R^2}.
\end{align}

Only the leading term of shortwave approximation up to 0PN order \(A^0= Q/R'\) is needed for this result, where $A^0 = A^0_\mathcal{M} + A^0_\mathcal{W}$ denotes the total electromagnetic wave-zone solution. We emphasize that $A^0$ here represents the scalar effective source generating the scalar wave-zone solution instead of the electromagnetic wave-zone solution $A^0_\mathcal{W}$ from wave-zone contributions. The next-to-leading term in \(A^0\), proportional to \(R'^{-2}\), generates the wave-zone contribution that starts at order \(R^{-3}\). Therefore it can be neglected at the present accuracy. Since the wave-zone scalar contribution starts at \(R^{-2}\), it does not affect the angular momentum flux calculation, which is determined by the radiative \(R^{-1}\) field.

The wave-zone contribution to the electromagnetic potential can be evaluated in the
same way, following the wave-zone integration method of Refs.~\cite{Will:1996zj, Shiralilou:2021mfl}. The relevant wave-zone contributions are
\begin{align}
    A^0_\mathcal{W}
    &=
    -\frac{aGM(X_1\alpha_1+X_2\alpha_2) Q}{c^2R^2},
    \\
    A^j_\mathcal{W}
    &=
    -\frac{aGM(X_1\alpha_1+X_2\alpha_2) (X_2q_1-X_1q_2)}{2c^3R^2} \notag \\
    &\quad \times
    \left[
        v^j+N^j(\mathbf N\cdot\mathbf v)
    \right],
\end{align}
Both terms start at order \(R^{-2}\), and therefore do not modify the radiative \(R^{-1}\) electromagnetic field entering the angular momentum flux.

The wave-zone contribution to the spatial gravitational potential is also obtained by
the same method. The wave-zone contribution is
\begin{equation}
    h^{jk}_\mathcal{W}
    =
    \frac{G}{c^4R^2}
    \left[
        GM^2
        +
        GM^2(X_1\alpha_1+X_2\alpha_2)^2
        -
        Q^2
    \right]
    N^jN^k .
\end{equation}
This contribution is again of order \(R^{-2}\), and hence it gives no correction to the radiative \(R^{-1}\) gravitational waveform relevant for the angular momentum flux. And in the TT gauge, since $h^{jk}$ at order $R^{-2}$ consists of radial unit direction vector $N^j$ only, it vanishes directly:
\begin{equation}
    h^{jk}_{\mathcal{W},\text{TT}}
    =
    0
    +
    \mathcal{O}(R^{-3}).
\end{equation}
This implies that the wave-zone contribution can be neglected at the next-to-leading order \(R^{-2}\) in the TT gauge.

\section{Gravitational multipole moment and waveform coefficients} \label{app:grav_moment_coeff}

In this appendix, we show the coefficients appearing in the mass multipole moments and waveform.

In Sec.~\ref{sec:grav_field}, the mass octupole moment $I^{jk}_l$ is given by
\begin{align}
    I^{jk}_l &= I^{jk}_{l,V,m} + I^{jk}_{l,V,\varphi} + I^{jk}_{l,V,A} + I^{jk}_{l,V,\text{LL}} + \mathcal{O}(c^{-4}),
\end{align}
\begin{widetext}
with the coefficients
    \begin{align}
        I^{jk}_{l,V,m} &= \frac{1}{2} \frac{d}{dt} \sum_A m_A \left( v_A^j x_A^k x_A^l + v_A^k x_A^j x_A^l - v_A^l x_A^j x_A^k \right) \left[ 1 + \frac{v_A^2}{2c^2} + \sum_{B \neq A} \frac{G m_B}{c^2 r_{AB}} (3 - \alpha_A \alpha_B) \right], \\
        I^{jk}_{l,V,\varphi} &= \frac{G}{2 c^2} \frac{d}{dt} \sum_A \sum_{B \neq A} m_A m_B \alpha_A \alpha_B v_A^a C_{ajkl}(A,B), \\
        I^{jk}_{l,V,A} &= -\frac{1}{2c^2} \frac{d}{dt} \sum_A \sum_{B \neq A} q_A q_B \Big[ v_A^j D_{iikl}(A,B) + v_A^k D_{iijl}(A,B) - v_A^l D_{iijk}(A,B) - v_B^a C_{ajkl}(A,B) \Big], \\
        I^{jk}_{l,V,\text{LL}} &= \frac{G}{2c^2} \frac{d}{dt} \sum_A \sum_{B \neq A} m_A m_B \Big[ -3 v_A^a C_{ajkl}(A,B) + 4 v_B^a C_{ajkl}(A,B) \notag \\
        &\quad - 4 \left( v_A^j D_{iikl}(A,B) + v_A^k D_{iijl}(A,B) - v_A^l D_{iijk}(A,B) \right) \Big],
    \end{align}
\end{widetext}
where $C_{ajkl}(A,B)$ and $D_{ajkl}(A,B)$ as results of integration read
\begin{widetext}
    \begin{align}
        C_{ajkl}(A,B) &= D_{ajkl}(A,B) + D_{akjl}(A,B) - D_{aljk}(A,B) \notag \\
        &= \frac{r_{AB}}{6} \Big[ (\delta_a^j n_{AB}^k n_{AB}^l + \delta_a^k n_{AB}^j n_{AB}^l + \delta_a^l n_{AB}^j n_{AB}^k) - n_{AB}^a n_{AB}^j n_{AB}^k n_{AB}^l \notag \\
        &\quad - \frac{1}{2} n_{AB}^a (\delta_{jk} n_{AB}^l + \delta_{jl} n_{AB}^k + \delta_{kl} n_{AB}^j) - \frac{1}{2} (\delta_{jk} \delta_a^l + \delta_{jl} \delta_a^k + \delta_{kl} \delta_a^j) \Big] \notag \\
        &\quad + \frac{x_B^l}{2} \Big[ \delta_a^j n_{AB}^k + \delta_a^k n_{AB}^j - n_{AB}^a n_{AB}^j n_{AB}^k - n_{AB}^a \delta_{jk} \Big] \notag \\
        &\quad + \frac{1}{2 r_{AB}} \Big[ x_B^k x_B^l (\delta_a^j - n_{AB}^a n_{AB}^j) + x_B^j x_B^l (\delta_a^k - n_{AB}^a n_{AB}^k) - x_B^j x_B^k (\delta_a^l - n_{AB}^a n_{AB}^l) \Big], \\
        D_{ajkl}(A,B) &= \frac{r_{AB}}{6} \Big[ (\delta_a^j n_{AB}^k n_{AB}^l + \delta_a^k n_{AB}^j n_{AB}^l + \delta_a^l n_{AB}^j n_{AB}^k) - n_{AB}^a n_{AB}^j n_{AB}^k n_{AB}^l \notag \\
        &\quad - \frac{1}{2} n_{AB}^a (\delta_{jk} n_{AB}^l + \delta_{jl} n_{AB}^k + \delta_{kl} n_{AB}^j) - \frac{1}{2} (\delta_{jk} \delta_a^l + \delta_{jl} \delta_a^k + \delta_{kl} \delta_a^j) \Big] \notag \\
        &\quad + \frac{x_B^l}{4} \Big[ \delta_a^j n_{AB}^k + \delta_a^k n_{AB}^j - n_{AB}^a n_{AB}^j n_{AB}^k - n_{AB}^a \delta_{jk} \Big] + \frac{x_B^k}{4} \Big[ \delta_a^j n_{AB}^l + \delta_a^l n_{AB}^j - n_{AB}^a n_{AB}^j n_{AB}^l - n_{AB}^a \delta_{jl} \Big] \notag \\
        &\quad + \frac{x_B^k x_B^l}{2 r_{AB}} (\delta_a^j - n_{AB}^a n_{AB}^j).
    \end{align}
\end{widetext}

For the mass 32-pole moment $I^{jk}_{lmn}$, the coefficient $E_{ajklm}(A,B)$ is given by
\begin{widetext}
    \begin{align}
        E_{ajklm}(A,B)
&= \frac{r_{AB}^2}{8} \Big[
\delta_a^j n_{AB}^k n_{AB}^l n_{AB}^m
+ \delta_a^k n_{AB}^j n_{AB}^l n_{AB}^m
+ \delta_a^l n_{AB}^j n_{AB}^k n_{AB}^m
+ \delta_a^m n_{AB}^j n_{AB}^k n_{AB}^l
- n_{AB}^a n_{AB}^j n_{AB}^k n_{AB}^l n_{AB}^m
\Big] \notag \\
&\quad - \frac{r_{AB}^2}{24} \Big\{
\left(
\delta_a^j n_{AB}^k
+ \delta_a^k n_{AB}^j
+ n_{AB}^a n_{AB}^j n_{AB}^k
\right)\delta_{lm}
+ \left(
\delta_a^j n_{AB}^l
+ \delta_a^l n_{AB}^j
+ n_{AB}^a n_{AB}^j n_{AB}^l
\right)\delta_{km} \notag \\
&\quad + \left(
\delta_a^j n_{AB}^m
+ \delta_a^m n_{AB}^j
+ n_{AB}^a n_{AB}^j n_{AB}^m
\right)\delta_{kl}
+ \left(
\delta_a^k n_{AB}^l
+ \delta_a^l n_{AB}^k
+ n_{AB}^a n_{AB}^k n_{AB}^l
\right)\delta_{jm} \notag \\
&\quad + \left(
\delta_a^k n_{AB}^m
+ \delta_a^m n_{AB}^k
+ n_{AB}^a n_{AB}^k n_{AB}^m
\right)\delta_{jl}
+ \left(
\delta_a^l n_{AB}^m
+ \delta_a^m n_{AB}^l
+ n_{AB}^a n_{AB}^l n_{AB}^m
\right)\delta_{jk}
\Big\} \notag \\
&\quad + \frac{r_{AB}^2}{24} n_{AB}^a
\left(
\delta_{jk}\delta_{lm}
+ \delta_{jl}\delta_{km}
+ \delta_{jm}\delta_{kl}
\right) \notag \\
&\quad + \frac{x_B^k r_{AB}}{6} \Big[
\delta_a^j n_{AB}^l n_{AB}^m
+ \delta_a^l n_{AB}^j n_{AB}^m
+ \delta_a^m n_{AB}^j n_{AB}^l
- n_{AB}^a n_{AB}^j n_{AB}^l n_{AB}^m \notag \\
&\quad - \frac{1}{2} n_{AB}^a
\left(
\delta_{jl} n_{AB}^m
+ \delta_{jm} n_{AB}^l
+ \delta_{lm} n_{AB}^j
\right)
- \frac{1}{2}
\left(
\delta_{jl} \delta_a^m
+ \delta_{jm} \delta_a^l
+ \delta_{lm} \delta_a^j
\right)
\Big] \notag \\
&\quad + \frac{x_B^l r_{AB}}{6} \Big[
\delta_a^j n_{AB}^k n_{AB}^m
+ \delta_a^k n_{AB}^j n_{AB}^m
+ \delta_a^m n_{AB}^j n_{AB}^k
- n_{AB}^a n_{AB}^j n_{AB}^k n_{AB}^m \notag \\
&\quad - \frac{1}{2} n_{AB}^a
\left(
\delta_{jk} n_{AB}^m
+ \delta_{jm} n_{AB}^k
+ \delta_{km} n_{AB}^j
\right)
- \frac{1}{2}
\left(
\delta_{jk} \delta_a^m
+ \delta_{jm} \delta_a^k
+ \delta_{km} \delta_a^j
\right)
\Big] \notag \\
&\quad + \frac{x_B^m r_{AB}}{6} \Big[
\delta_a^j n_{AB}^k n_{AB}^l
+ \delta_a^k n_{AB}^j n_{AB}^l
+ \delta_a^l n_{AB}^j n_{AB}^k
- n_{AB}^a n_{AB}^j n_{AB}^k n_{AB}^l \notag \\
&\quad - \frac{1}{2} n_{AB}^a
\left(
\delta_{jk} n_{AB}^l
+ \delta_{jl} n_{AB}^k
+ \delta_{kl} n_{AB}^j
\right)
- \frac{1}{2}
\left(
\delta_{jk} \delta_a^l
+ \delta_{jl} \delta_a^k
+ \delta_{kl} \delta_a^j
\right)
\Big] \notag \\
&\quad + \frac{x_B^k x_B^l}{4} \Big[
\delta_a^j n_{AB}^m
+ \delta_a^m n_{AB}^j
- n_{AB}^a n_{AB}^j n_{AB}^m
- n_{AB}^a \delta_{jm}
\Big] \notag \\
&\quad + \frac{x_B^k x_B^m}{4} \Big[
\delta_a^j n_{AB}^l
+ \delta_a^l n_{AB}^j
- n_{AB}^a n_{AB}^j n_{AB}^l
- n_{AB}^a \delta_{jl}
\Big] \notag \\
&\quad + \frac{x_B^l x_B^m}{4} \Big[
\delta_a^j n_{AB}^k
+ \delta_a^k n_{AB}^j
- n_{AB}^a n_{AB}^j n_{AB}^k
- n_{AB}^a \delta_{jk}
\Big] \notag \\
&\quad + \frac{x_B^k x_B^l x_B^m}{2r}
\left(
\delta_a^j
- n_{AB}^a n_{AB}^j
\right).
    \end{align}
\end{widetext}

In Sec.~\ref{sec:grav_waveform}, the 1.5PN contribution $h^{jk}_{R^{-1},1.5,\text{TT}}$ at leading order $\mathcal{O}(R^{-1})$ is given by
\begin{widetext}
    \begin{align}
    h^{jk}_{R^{-1},1.5,\text{TT}}
&=
\Big\{
\mathcal H_{rr}\,r^jr^k
+\mathcal H_{rv}\left(r^jv^k+v^jr^k\right)
+\mathcal H_{vv}\,v^jv^k
\Big\}_{\mathrm{TT}},
\end{align}
with the coefficients
\begin{align}
\mathcal H_{rr}
&=
(X_1-X_2)\frac{\eta G_{12}M}{r^3}
\Bigg\{
-\frac{-1+2\nu}{4}
\left(\mathbf N\cdot\mathbf v\right)^3
-\frac{5(-1+2\nu)\dot r}{8r^3}
\left(
6\frac{G_{12}M}{r}
-7\dot r^2
+3v^2
\right)
\left(\mathbf N\cdot\mathbf r\right)^3
\notag\\
&\quad
+\frac{-1+2\nu}{8r^2}
\left(
28\frac{G_{12}M}{r}
+9(-5\dot r^2+v^2)
\right)
\left(\mathbf N\cdot\mathbf r\right)^2
\left(\mathbf N\cdot\mathbf v\right)
\notag\\
&\quad
-\frac{\left(\mathbf N\cdot\mathbf v\right)}{24}
\Bigg[
\frac{2q_1q_2}{\eta G_{12}M}
\left(
4\frac{G_{12}M}{r}
+3(-1+3\nu)(3\dot r^2-v^2)
\right)
-8(1+\alpha_1\alpha_2)\frac{G}{G_{12}}
\frac{G_{12}M}{r}(-2+3\nu)
\notag\\
&\qquad
+6\frac{G}{G_{12}}
\left(5+3\alpha_1\alpha_2(-1+\nu)+3\nu\right)
(-3\dot r^2+v^2)
\notag\\
&\qquad
+4\frac{G_{12}M}{r}(1-3\nu+6\Gamma)
+3\left[
(1+2\nu-6\Gamma)v^2
+3(-1+4\Gamma)\dot r^2
\right]
+12\frac{G^2}{G_{12}^2}
\frac{G_{12}M}{r}\Pi
\Bigg]
\notag\\
&\quad
+\left(\mathbf N\cdot\mathbf r\right)
\Bigg[
\frac{9(-1+2\nu)\dot r}{4r}
\left(\mathbf N\cdot\mathbf v\right)^2
+\frac{\dot r}{24r}
\Bigg[
\frac{2q_1q_2}{\eta G_{12}M}
\left(
2(8-9\nu)\frac{G_{12}M}{r}
+3(-1+3\nu)(5\dot r^2-3v^2)
\right)
\notag\\
&\qquad\qquad
-2(1+\alpha_1\alpha_2)\frac{G}{G_{12}}
\left(
10(-2+3\nu)\frac{G_{12}M}{r}
+3(-1+3\nu)(5\dot r^2-3v^2)
\right)
\notag\\
&\qquad\qquad
-15\dot r^2
+9v^2(1+2\nu-2\Gamma)
+2(5-6\nu+24\Gamma)\frac{G_{12}M}{r}
+48\frac{G^2}{G_{12}^2}
\frac{G_{12}M}{r}\Pi
\Bigg]
\Bigg]
\Bigg\},
\\[1ex]
\mathcal H_{rv}
&=
(X_1-X_2)\frac{\eta G_{12}M}{r^2}
\Bigg\{
\frac{27(-1+2\nu)\dot r}{4r^2}
\left(\mathbf N\cdot\mathbf r\right)^2
\left(\mathbf N\cdot\mathbf v\right)
\notag\\
&\quad
-\frac{\dot r}{12}
\Bigg[
3\Bigg(
1+2\nu
+2\frac{G}{G_{12}}
\big(3+\alpha_1\alpha_2(-1+\nu)+\nu\big)
\Bigg)
+\frac{2(2-3\nu)q_1q_2}{\eta G_{12}M}
\Bigg]
\left(\mathbf N\cdot\mathbf v\right)
\notag\\
&\quad
+\frac{(-1+2\nu)}{24r^3}
\left(
44\frac{G_{12}M}{r}
+21(-5\dot r^2+v^2)
\right)
\left(\mathbf N\cdot\mathbf r\right)^3
\notag\\
&\quad
+\left(\mathbf N\cdot\mathbf r\right)
\Bigg[
-\frac{15(-1+2\nu)}{4r}
\left(\mathbf N\cdot\mathbf v\right)^2
-\frac{1}{24r}
\Bigg[
-\frac{2(-2+3\nu)q_1q_2}{\eta G_{12}M}
\left(
4\frac{G_{12}M}{r}
-3\dot r^2+v^2
\right)
\notag\\
&\qquad\qquad
-8(1+\alpha_1\alpha_2)\frac{G}{G_{12}}
\frac{G_{12}M}{r}(-2+3\nu)
+6\frac{G}{G_{12}}
\big(-3+\nu+\alpha_1\alpha_2(1+\nu)\big)
(-3\dot r^2+v^2)
\notag\\
&\qquad\qquad
+4\frac{G_{12}M}{r}(1-9\nu+6\Gamma)
+3\Big[
(1+6\nu-2\Gamma)v^2
-3(1+4\Gamma)\dot r^2
\Big]
+36\frac{G^2}{G_{12}^2}
\frac{G_{12}M}{r}\Pi
\Bigg]
\Bigg]
\Bigg\},
\\[1ex]
\mathcal H_{vv}
&=
(X_1-X_2)\eta
\Bigg\{
(-1+2\nu)
\left(\mathbf N\cdot\mathbf v\right)^3
-\frac{29(-1+2\nu)}{4r^2}
\frac{G_{12}M}{r}
\left(\mathbf N\cdot\mathbf r\right)^2
\left(\mathbf N\cdot\mathbf v\right)
+\frac{17(-1+2\nu)\dot r}{4r^3}
\frac{G_{12}M}{r}
\left(\mathbf N\cdot\mathbf r\right)^3
\notag\\
&\quad
+\frac{\left(\mathbf N\cdot\mathbf v\right)}{4}
\Bigg[
\frac{G_{12}M}{r}
\Bigg(
1
-2(1+\alpha_1\alpha_2)\frac{G}{G_{12}}(-1+\nu)
-6\nu
-\frac{2(-1+\nu)q_1q_2}{\eta G_{12}M}
\Bigg)
+2(-1+5\nu)v^2
\Bigg]
\notag\\
&\quad
+\frac{\dot r\left(\mathbf N\cdot\mathbf r\right)}{4r}
\frac{G_{12}M}{r}
\Bigg[
\frac{2(-1+\nu)q_1q_2}{\eta G_{12}M}
+2\frac{G}{G_{12}}
\big(
5+\alpha_1\alpha_2(-3+\nu)+\nu
\big)
-5+6\nu-12\Gamma
\Bigg]
\Bigg\}.
    \end{align}
And the 1PN contribution $h^{jk}_{R^{-2},1,\text{TT}}$ at next-to-leading order $\mathcal{O}(R^{-2})$ is given by
    \begin{align}
    h^{jk}_{R^{-2},1,\text{TT}}
&=
\Bigg\{
\mathcal H^{(2)}_{rr}\,r^jr^k
+\mathcal H^{(2)}_{rv}\left(r^jv^k+v^jr^k\right)
+\mathcal H^{(2)}_{vv}\,v^jv^k
\Bigg\}_{\text{TT}},
\end{align}
with the coefficients
\begin{align}
\mathcal H^{(2)}_{rr}
&=
(X_1-X_2)\frac{\eta G_{12}M}{r^2}
\Bigg\{
\frac{9(-1+2\nu)\dot r}{2r^2}
(\mathbf N\cdot\mathbf r)^2(\mathbf N\cdot\mathbf v)
+\frac{-1+2\nu}{4r^3}
\left[
10\frac{G_{12}M}{r}
+3(-5\dot r^2+v^2)
\right]
(\mathbf N\cdot\mathbf r)^3
\notag\\
&\quad
-\frac{\dot r}{6}
\Bigg[
(1-3\nu)\frac{q_1q_2}{\eta G_{12}M}
+1
+\frac{G}{G_{12}}
\big(8+3\nu+\alpha_1\alpha_2(-4+3\nu)\big)
+3(\nu-\Gamma)
\Bigg]
(\mathbf N\cdot\mathbf v)
\notag\\
&\quad
+\frac{(\mathbf N\cdot\mathbf r)}{r}
\Bigg[
-\frac{3(-1+2\nu)}{2}
(\mathbf N\cdot\mathbf v)^2
-\frac{1}{12}
\Bigg[
\frac{q_1q_2}{\eta G_{12}M}
\left(
(4-6\nu)\frac{G_{12}M}{r}
+(-1+3\nu)(3\dot r^2-v^2)
\right)
\notag\\
&\qquad\qquad
-(1+\alpha_1\alpha_2)\frac{G}{G_{12}}
\left(
2(-2+3\nu)\frac{G_{12}M}{r}
+(-1+3\nu)(3\dot r^2-v^2)
\right)
\notag\\
&\qquad\qquad
-3(1+3\nu)\dot r^2
+(1+6\nu-3\Gamma)v^2
+2(-5+12\nu+3\Gamma)\frac{G_{12}M}{r}
+6\frac{G^2}{G_{12}^2}\frac{G_{12}M}{r}\Pi
\Bigg]
\Bigg]
\Bigg\},
\\[1ex]
\mathcal H^{(2)}_{rv}
&=
(X_1-X_2)\frac{\eta G_{12}M}{r}
\Bigg\{
-\frac{15(-1+2\nu)}{2r^2}
(\mathbf N\cdot\mathbf r)^2(\mathbf N\cdot\mathbf v)
-\frac{1}{12}
\Bigg[
2\left(
(-5+\alpha_1\alpha_2)\frac{G}{G_{12}}
+5(1-3\nu)
\right)
-\frac{q_1q_2}{\eta G_{12}M}
\Bigg]
(\mathbf N\cdot\mathbf v)
\notag\\
&\quad
+\frac{9(-1+2\nu)\dot r}{2r^3}
(\mathbf N\cdot\mathbf r)^3
-\frac{\dot r}{12r}
\Bigg[
\frac{q_1q_2}{\eta G_{12}M}
+2\left(
2(-2+\alpha_1\alpha_2)\frac{G}{G_{12}}
+4+3\Gamma
\right)
\Bigg]
(\mathbf N\cdot\mathbf r)
\Bigg\},
\\[1ex]
\mathcal H^{(2)}_{vv}
&=
(X_1-X_2)\eta
\Bigg\{
6(-1+2\nu)
(\mathbf N\cdot\mathbf r)(\mathbf N\cdot\mathbf v)^2
-\frac{11(-1+2\nu)}{2r^2}
\frac{G_{12}M}{r}
(\mathbf N\cdot\mathbf r)^3
-2(-1+2\nu)r\dot r
(\mathbf N\cdot\mathbf v)
\notag\\
&\quad
-\frac{\mathbf N\cdot\mathbf r}{6}
\Bigg[
2(7-12\nu)\frac{G_{12}M}{r}
+\frac{G}{G_{12}}
\big(7+3\nu+\alpha_1\alpha_2(-5+3\nu)\big)
\frac{G_{12}M}{r}
+(-2+3\nu)\frac{q_1q_2}{\eta G_{12}M}
\frac{G_{12}M}{r}
-3(1+\nu)v^2
\Bigg]
\Bigg\}.
    \end{align}
\end{widetext}

\section{Angular momentum flux} \label{app:ang_moment_flux}

In this appendix we derive the angular momentum fluxes carried by the scalar,
electromagnetic, and gravitational radiation. The antisymmetric angular-momentum
flux tensor through a sphere of radius \(R\) is defined by
\begin{equation}
    \mathcal{J}^{jk}
    =
    R^2
    \int
    \left(
        x^j T_{X}^{kl}
        -
        x^k T_{X}^{jl}
    \right)
    N_l\,d\Omega ,\quad X=\varphi,A, \text{LL},
    \label{eq:J_general_definition}
\end{equation}
where \(\mathbf{N} = \mathbf{x}/R\). The total flux can be decomposed into scalar,
electromagnetic, and gravitational parts,
\begin{equation}
    \mathcal{J}^{jk}
    =
    \mathcal{J}^{jk}_{\varphi}
    +
    \mathcal{J}^{jk}_{A}
    +
    \mathcal{J}^{jk}_{\text{LL}}.
\end{equation}
Throughout this appendix, $\partial_\tau$ or a dot denotes a derivative with respect to the retarded
time \(\tau=t-R/c\). We also decompose the spatial derivative as
\begin{equation} 
    \partial_j
    =
    -\frac{N_j}{c}\partial_\tau
    +
    N_j\eth_R
    +
    \eth_j^T ,
    \label{eq:eth_derivative_definition}
\end{equation}
where
\begin{equation}
    \eth_R = N^j \eth_j, \qquad \eth^T_j = P_j^k \eth_k.
\end{equation}

The scalar field stress tensor in the wave zone is
\begin{equation}
    T^{jk}_{\varphi}
    =
    \frac{c^4}{4\pi G}
    \left[
        \partial^j\varphi\,\partial^k\varphi
        -
        \frac{1}{2}\delta^{jk}
        (\partial_i\varphi)(\partial_i\varphi)
    \right] .
\end{equation}
Substituting this expression into Eq.~\eqref{eq:J_general_definition}, the trace
term proportional to \(\delta^{jk}\) drops out because of antisymmetry.

Using Eq.~\eqref{eq:eth_derivative_definition}, we obtain
\begin{equation}
    N_l\partial^l\varphi
    =
    -\frac{1}{c}\dot\varphi
    +
    \mathcal{O}(R^{-2})
\end{equation}
and
\begin{equation}
    x^j\partial^k\varphi-x^k\partial^j\varphi
    =
    R
    \left(
        N^j\partial_T^k\varphi
        -
        N^k\partial_T^j\varphi
    \right),
\end{equation}
we obtain the scalar angular momentum flux
\begin{equation}
    \mathcal{J}^{jk}_{\varphi}
    =
    -\frac{c^3}{4\pi G}
    \int
    R^3\dot{\varphi}
    \left(
        N^j\partial_T^k\varphi
        -
        N^k\partial_T^j\varphi
    \right)d\Omega .
    \label{eq:J_scalar_final}
\end{equation}
Only the angular dependence of the radiative \(R^{-1}\) scalar field contributes
to the finite flux at infinity.

For the electromagnetic field and gravitational field, the angular momentum $\mathcal{J}^{jk}_{A}$ and $\mathcal{J}^{jk}_{\text{LL}}$ can be computed in the same method:
\begin{widetext}
    \begin{align}
    \mathcal{J}^{jk}_A &= \frac{1}{4 \pi c} \int \Bigg\{ R^3 \Bigg[ N^j \Big( \dot{A}^0 \eth^k_T A^0 - \dot{A}^l \eth^k_T A_l + \dot{A}^l \eth^T_l A^k - \dot{A}^k \eth^T_l A^l - h^{00} \dot{A}^k N_l \dot{A}^l \Big) \notag \\
    &\quad\ - N^k \Big( \dot{A}^0 \eth^j_T A^0 - \dot{A}^l \eth^j_T A_l + \dot{A}^l \eth^T_l A^j - \dot{A}^j \eth^T_l A^l - h^{00} \dot{A}^j N_l \dot{A}^l \Big) \Bigg] \notag \\
    & + R^2 \Bigg[ 2(N_l A^l) (N^j \dot{A}^k - N^k \dot{A}^j) - (N_l \dot{A}^l) (N^j A^k - N^k A^j) \Bigg] \Bigg\} d\Omega, \\
    \mathcal{J}^{jk}_{\text{LL}} &= \frac{c^3}{16\pi G} \int R^2 \Bigg[ h_{\text{TT}}^{jp} \dot{h}_{\text{TT}}^{kp} - h_{\text{TT}}^{kp} \dot{h}_{\text{TT}}^{jp} - \frac{1}{2} \dot{h}_{\text{TT}}^{pq} (x^j \partial^k - x^k \partial^j) h^{\text{TT}}_{pq} \Bigg] d\Omega.
    \end{align}
\end{widetext}

\bibliography{references} 
\bibliographystyle{apsrev4-2}
\end{document}